\documentclass[]{agujournal2018}
\usepackage{amsmath,amssymb}
\usepackage{apacite}
\usepackage{url} 
\draftfalse
\journalname{JGR-Space Physics}

\newcommand{\ignore}[1]{}  
\newcommand{\todo}[1]{}
\newcommand{\note}[2]{}
\newcommand\tocite[1]{}
\newcommand{\V}[1]{\mathbf{#1}} 
\newcommand{\gkeyll}{{\tt Gkeyll}}

\begin{document}

\title{An extended MHD study of the 16 October 2015 MMS diffusion region crossing}

\authors{J. M. TenBarge\affil{1,2}, J. Ng\affil{1,2,3}, J. Juno\affil{4},  L. Wang\affil{1,2}, A. H. Hakim\affil{2,5}, and A. Bhattacharjee\affil{1,2,5}}

\affiliation{1}{Department of Astrophysical Sciences, Princeton University, Princeton, NJ, USA.}
\affiliation{2}{Princeton Center for Heliophysics, Princeton Plasma Physics Laboratory, Princeton, NJ, USA.}
\affiliation{3}{Department of Astronomy, University of Maryland, College Park, MD, USA.}
\affiliation{4}{IREAP, University of Maryland, College Park, MD, USA.}
\affiliation{5}{Princeton Plasma Physics Laboratory, Princeton, NJ, USA.}

\correspondingauthor{J. M. TenBarge}{tenbarge@princeton.edu}

\begin{keypoints}
\item An extended MHD model evolving the full electron pressure tensor and capturing aspects of Landau damping is applied to the MMS Burch event

\item Many aspects of the fluid model agree well with in situ observations, including strong parallel electron heating near the current layer

\item The event study demonstrates the efficacy of the ten moment model for global magnetosphere modeling efforts
\end{keypoints}

\begin{abstract}
The Magnetospheric Multiscale (MMS) mission has given us unprecedented access to high cadence particle and field data of magnetic reconnection at Earth's magnetopause. MMS first passed very near an X-line on 16 October 2015, the Burch event, and has since observed multiple X-line crossings. Subsequent 3D particle-in-cell (PIC) modeling efforts of and comparison with the Burch event have revealed a host of novel physical insights concerning magnetic reconnection, turbulence induced particle mixing, and secondary instabilities. In this study, we employ the \gkeyll~simulation framework to study the Burch event with different classes of extended, multi-fluid magnetohydrodynamics (MHD), including  models that incorporate important kinetic effects, such as the electron pressure tensor, with physics-based closure relations designed to capture linear Landau damping. Such fluid modeling approaches are able to capture different levels of kinetic physics in global simulations and are generally less costly than fully kinetic PIC. We focus on the additional physics one can capture with increasing levels of fluid closure refinement via comparison with MMS data and existing PIC simulations.
\end{abstract}

\section{Introduction}\label{sec:intro}
The coupled sun-Earth system is a critically important problem for protecting mankind's technological infrastructure from damaging space weather events, but planetary magnetospheres represent a multi-scale, multi-process physical system that is weakly collisional and coupled to the highly dynamic sun, placing great demand on modeling efforts. Modern \textit{in situ} spacecraft missions like the Magnetospheric Multiscale (MMS) mission \citep{Burch:2016b} are providing field and particle data at unprecedented temporal and spatial resolution, revolutionizing our understanding of the fundamental plasma processes occurring in Earth's magnetosphere, such as magnetic reconnection, shocks, and turbulence. Magnetic reconnection of the shocked plasma at Earth's magnetosheath transports energy from the solar wind plasma into Earth's magnetosphere \citep{Paschmann:1979} and is a core process for understanding the sun-Earth coupling.


Despite the impressive quality of \textit{in situ} measurements, numerical modelling remains our best resource for understanding the global behavior of planetary magnetospheres. Today, the three main numerical approaches are: 1) fully kinetic models such as particle-in-cell (PIC) and continuum Vlasov; 2) hybrid models wherein one species, typically ions, is treated kinetically and the other as a fluid; and 3) fluid models, such as magnetohydrodynamics (MHD), Hall MHD, and various extensions of MHD. Each of these modeling approaches has inherent advantages. For example, fully kinetic models are capable of capturing all of the fundamental plasma physics processes occurring in a magnetosphere; however, such complete modeling comes at the expense of computational complexity, making global simulations of Earth's magnetosphere too costly to perform. Hybrid approaches are attractive because of their reduced computational cost relative to fully kinetic models, but the electron are typically treated as a massless, isothermal fluid, which is problematic for sub-ion physics such as reconnection, which will be diffused by explicit or numerical resistivity at the simulation grid scale. Additionally, one may be given to assume that since the ions are kinetic, they are treated fully; however, the kinetic ions are coupled to fluid electrons through Maxwell's equations, which can have consequences at large-scales, such as fast magnetosonic modes being undamped \citep{Parashar:2014,Groselj:2017}. Fluid simulations are the most commonly employed approach to model global magnetospheres due to their significantly reduced computational demand, but they assume that the plasma is highly collisional and thus near thermodynamic equilibrium; neither is a good assumption in the weakly collisional plasmas that permeate the solar wind and magnetospheres. Fluid models can be improved by retaining progressively higher fluid moments, with each additional moment retained by the closure corresponding to the fluid model more accurately representing the fully kinetic system. The closure employed by the fluid model can also be tuned to better approximate kinetic effects.


Magnetic reconnection plays a fundamental role in coupling the sun-Earth system; therefore, any fluid model employed in global magnetosphere modeling should minimally capture the physics of reconnection accurately. Reconnection is ultimately controlled by electron demagnetization, which is governed by a combination of electron inertia and pressure agyrotropy \citep{Vasyliunas:1975}. Based upon this condition, the minimum requirement for a fluid model for global modeling is that the electron closure retain finite mass and evolve the full pressure tensor, implying at least ten moments be retained for each species, $n$, $\V{u}$, and $\mathcal{P}_{ij}$, i.e., $1+3+6$ moments. Since the plasmas of interest are weakly collisional, the closure for a ten moment model should incorporate collisionless effects, such as Landau damping. The Hammett-Perkins \cite{Hammett:1990} approach to closure is one such attempt to incorporate aspects of linear Landau damping in a fluid model. The Hammett-Perkins closure is based upon a linearization of the Vlasov equation, followed by integration over velocity space to obtain an expression for the unclosed fluid moment. 

The \gkeyll~simulation framework includes both a two-fluid, five-moment model that achieves closure via truncation of moments beyond the isotropic pressure, $p$, and a ten moment model with an approximation of the Hammett-Perkins closure for the heat flux. The ten moment model has been well compared to PIC in the anti-parallel reconnection case \citep{Wang:2015} and island coalescence \citep{Ng:2015}. \gkeyll~has also been successfully employed to model the magnetosphere of the Jovian moon Ganymede \citep{Wang:2018}. To further evaluate the efficacy of the ten moment model, we consider here an asymmetric reconnection event based upon the 16 October 2015 MMS magnetosheath crossing \citep{Burch:2016}. The large asymmetry of this event provides a stringent test for the ten moment model in the two and three dimensions that can be validated against \textit{in situ} observations and existing PIC simulations of the event.

The paper is organized as follows. In section~\ref{sec:models}, we introduce the \gkeyll~framework and discuss the closures for the two fluid five ($n$, $\V{u}$, and $p$) and ten moment ($n$, $\V{u}$, and $\mathcal{P}_{ij}$) models that are part of \gkeyll. Section~\ref{sec:ics} presents the initial conditions derived from the Burch event and the closure parameters chosen for the ten moment model. Simulation results for the ten and five moment models are provided in section~\ref{sec:results}. Section~\ref{sec:discussion} places the results of the paper in the context of contemporary simulation methods. Finally, in section~\ref{sec:conclusions}, we present a summary of this paper.

\section{The Five- and Ten-Moment Models}\label{sec:models}
The simulation framework we will use for this study is \gkeyll, which includes solvers for continuum Vlasov-Maxwell \cite{Juno:2018}, gyrokinetics \cite{Shi:2015}, and five and ten moment two fluid equations \cite{Hakim:2006,Hakim:2008,Wang:2015}. In this work, we focus on the five- and ten-moment two-fluid models. 

The fluid models evolve the electromagnetic fields through Maxwell's equations
\begin{equation}
    \nabla \times \V{E} = -\frac{\partial \V{B}}{\partial t},
\end{equation}
\begin{equation}
    \nabla \times \V{B} = \mu_0 \V{J} + \frac{1}{\mu_0 \epsilon_0} \frac{\partial \V{E}}{\partial t},
\end{equation}
where $\V{E}$ and $\V{B}$ are the electric and magnetic field, $\mu_0$ and $\epsilon_0$ are the permeability and permittivity of free space, and $\V{J} = \sum_s n_s q_s \V{u}_s$, with $q_s$, $n_s$, and $\V{u}_s$ representing the species charge, density, and velocity. In the following, we drop the species subscripts for notational elegance. 

The evolution equations for the moments are obtained via the standard procedure of taking moments of the Vlasov equation:
\begin{equation}\label{eq:n}
    \frac{\partial n}{\partial t} + \frac{\partial n u_i}{\partial x_i} = 0,
\end{equation}
\begin{equation}\label{eq:u}
    m \frac{\partial n u_i}{\partial t} + \frac{\partial \mathcal{P}_{ij}}{\partial x_j}  = n q (E_i + \epsilon_{ijk} u_j B_k),
\end{equation}
\begin{equation}\label{eq:Pij}
    \frac{\partial \mathcal{P}_{ij}}{\partial t} + \frac{\partial \mathcal{Q}_{ijk}}{\partial x_k} = n q u_{[i}E_{j]} + \frac{q}{m} \epsilon_{[ikl}\mathcal{P}_{kj]}B_l,
\end{equation}
where $\epsilon_{ijk}$ is is the completely anti-symmetric Levi-Cevita tensor, which is defined to be $\pm 1$ for even/odd permutations and zero otherwise, and square brackets around indices represent the minimal sum over permutations of free indices needed to yield completely symmetric tensors.
\begin{equation}
    \mathcal{P}_{ij} = m\int v_i v_j f d\V{v} = P_{ij} + mnu_i u_j,
\end{equation}
is the pressure tensor and
\begin{equation}
    \mathcal{Q}_{ijk} = m\int v_i v_j v_k f d\V{v} = Q_{ijk} + u_{[i}\mathcal{P}_{jk]} - 2mnu_i u_j u_k
\end{equation}
is the heat-flux tensor, where $P_{ij}$ and $Q_{ijk}$ are the rest-frame pressure and heat-flux tensors. At this level, the fluid equations are exact but an endless hierarchy, with the $n$th moment requiring knowledge of the $n+1$st moment.

\subsection{The Five-Moment Model}\label{sec:modelsFive}
One approach to achieve closure is a simple truncation by neglecting the heat-flux and considering only isotropic pressures. We begin by taking the trace of equation~\eqref{eq:Pij} to obtain the exact energy equation
\begin{equation}\label{eq:Efull}
    \frac{\partial \mathcal{E}}{\partial t} + \frac{1}{2}\frac{\partial \mathcal{Q}_{iik}}{\partial x_k} = n q \V{u}\cdot\V{E},
\end{equation}
where
\begin{equation}
    \mathcal{E} = \frac{1}{2}\mathcal{P}_{ii} = \frac{3}{2}p + \frac{1}{2} m n \V{u}^2,
\end{equation}
$p = P_{ii}/3$,
\begin{equation}
    \frac{1}{2}\mathcal{Q}_{iik} = q_k + u_k (p + \mathcal{E}) + u_i \pi_{ik},
\end{equation}
$q_k = Q_{iik}/2$ is the heat-flux vector, $\pi_{ij} = P_{ij} - p \delta_{ij}$ is the viscous stress tensor, and $\delta_{ij}$ is the Kronecker delta function. Closure is obtained by setting $q_k = 0$ and $\pi_{ij} = 0$, reducing equation~\eqref{eq:Efull} to
\begin{equation}\label{eq:energy}
    \frac{\partial \mathcal{E}}{\partial t} + \frac{\partial}{\partial x_k} \left[u_k(p + \mathcal{E})\right] = n q \V{u}\cdot\V{E}.
\end{equation}
Thus, equations~\eqref{eq:n}-\eqref{eq:u} and \eqref{eq:energy} constitute a closed system of five-moment equations, $n$, $\V{u}$, and $p$, equations for each species. Note that this model reduces formally to Hall MHD in the limit $m_e \rightarrow 0$ and $\epsilon_0 \rightarrow 0$ \cite{Srinivasan:2011}.

\subsection{The Ten-Moment Model}\label{sec:modelsTen}
To retain additional kinetic physics relative to the five-moment model, we desire the full pressure tensor, thus constituting a ten moment model. Therefore, we must obtain an expression for $\partial Q_{ijk}/\partial x_k$ appearing implicitly in equation~\eqref{eq:Pij} to close the system. Additionally, since space plasmas tend to be weakly collisional, we seek a collisionless closure and thus choose to use a Hammett-Perkins \cite{Hammett:1990} inspired closure. The closure is based upon a linearization of the Vlasov equation, followed by integration over velocity space to obtain an expression for the unclosed fluid moment, which in this case is the heat flux. This approach leads to expressions involving plasma response functions in Fourier space, which are approximated by an $n$-pole Pad\'{e} approximant. Following \citet{Hammett:1990}, we choose to employ a $3$-pole approximation to the response function; however, since the approximation is in Fourier space, the closure is non-local along magnetic field lines. To reduce the computation demand of integrating along field lines and Fourier transforming, we further use a local approximation of the heat-flux divergence
\begin{equation}\label{eq:heatClosure}
    \frac{\partial Q_{ijk}}{\partial x_k} = v_t |k_0| (P_{ij} - p\delta_{ij}),
\end{equation}
where $v_t = \sqrt{T/m}$ is the thermal speed, and $k_0$ is a typical wavenumber, which defines a scale below which collisionless damping is thought to occur. Equation~\eqref{eq:heatClosure} together with equations~\eqref{eq:n}-\eqref{eq:Pij} represent the closed ten moment model for $n$, $\V{u}$, and $\mathcal{P}_{ij}$. Note that this closure closely resembles a collisional closure that scatters the system towards isotropy with collision frequency $\nu = v_t |k_0|$. $k_0 = 0$ would allow the pressure to evolve freely, while the limit $k_0 \rightarrow \infty$ isotropizes the pressure tensor at all scales, approaching the five-moment limit.

\section{Initial Conditions}\label{sec:ics}
The initial conditions for the study contained herein are motivated by the first observed Magnetospheric Multiscale (MMS) spacecraft crossing of a magnetic field X-line at Earth's magnetosheath \cite{Burch:2016}, hereafter referred to as the Burch event. The initial magnetic field and temperature profiles have the following form,
\begin{equation}\label{eq:profile}
    f(Q_s,Q_m) = 0.5 \left[ Q_s + Q_m + (Q_m - Q_s) \tanh{(y/w_0)}\right],
\end{equation}
where subscripts $s$ and $m$ correspond to magnetosheath and magnetosphere quantities, $w_0 = d_{is}$ is the initial current sheet width, $d_{is} = c / \omega_{pis}$ is the asymptotic ion inertial length on the sheath side, and $\omega_{pis}$ is the sheath ion plasma frequency. For notational simplicity, we will use $d_i \equiv d_{is}$ and $\Omega_{ci} = \Omega_{cis}$ going forward, where $\Omega_{cis}$ is the sheath gyrofrequency. The initial magnetic field profile is given by $B_x = f(B_{sx},B_{mx})$, where $B_{mx}/B_{sx} = -1.7$, with a uniform guide field $B_z/B_{sx} = 0.099$. The initial electron and ion temperature profiles are also given by equation~\eqref{eq:profile} and satisfy $T_{em}/T_{es} = 12.3$ and $T_{im}/T_{is} = 5.63$. The density profile is chosen to satisfy force balance across the sheet, which leads to $n_m/n_s = 16.7$. An initial vector potential perturbation of the form
\begin{equation}
    A_z = \psi_0 \cos{(2\pi x/ L_x)}\left[1-\cos{(4\pi y/L_y)}\right]
\end{equation}
is added to initiate reconnection, where $\psi_0 = 0.1$. The simulation dimensions are $(L_x,L_y) = (40.96,20.48)d_i$ in 2D and $(L_x,L_y,L_z) = (40.96,20.48,10.24)d_i$ in 3D, with fully periodic boundary conditions in all cases. The number of grid points are $(N_x,N_y) = (2048,1024)$ and $(N_x,N_y,N_z) = (1024,512,256)$ giving $\Delta x = 0.2 d_{es}$ and $\Delta x = 0.4 d_{es}$ in 2 and 3D. A reduced mass ratio of $m_i/m_e = 100$ and $c/v_{A\perp s} = 500$ are used in all cases, where $v_{A\perp s} = B_{sx}/\sqrt{\mu_0 m_i n_s}$ is the asymptotic sheath Alfv\'{e}n speed based upon the reconnecting field. To break the inherent symmetry of the initial conditions in 3D, noise with root-mean-square amplitude $B_{noise} = 0.0005 B_{sx}$ is added to the first 20 Fourier modes in the $x$, $y$, and $z$ directions of $B_x$ and $B_y$. We will refer to the 2D ten moment and 3D five and ten moment \gkeyll~simulations as 2D10, 3D5, and 3D10, respectively.

 Prior \gkeyll~reconnection studies \citep{Ng:2015,Wang:2015} have demonstrated that $k_{0s} d_{s} \sim 1$ is required to achieve agreement with particle-in-cell simulations, where subscript $s$ refers to species. However, a value of $1/10$ provides slightly better performance for turbulence simulations without significantly degrading the reconnection performance \citep{Juno:2018}. Since we expect these event simulations to be turbulent, we choose closure parameters for both ten moment model simulations to be $k_{0s} d_{s} = 1/10$. Formally, this choice for $k_0$ will permit pressure anisotropy and agyrotropy to develop to a scale $l_{0s} = 2\pi/k_{0s} \simeq d_s$ before the ten moment closure scatters the pressure toward isotropy.

\section{Results}\label{sec:results}
\subsection{Ten-Moment Simulation Results}\label{sec:resultsTen}
\begin{figure}[t]
\includegraphics[width=0.48\linewidth]{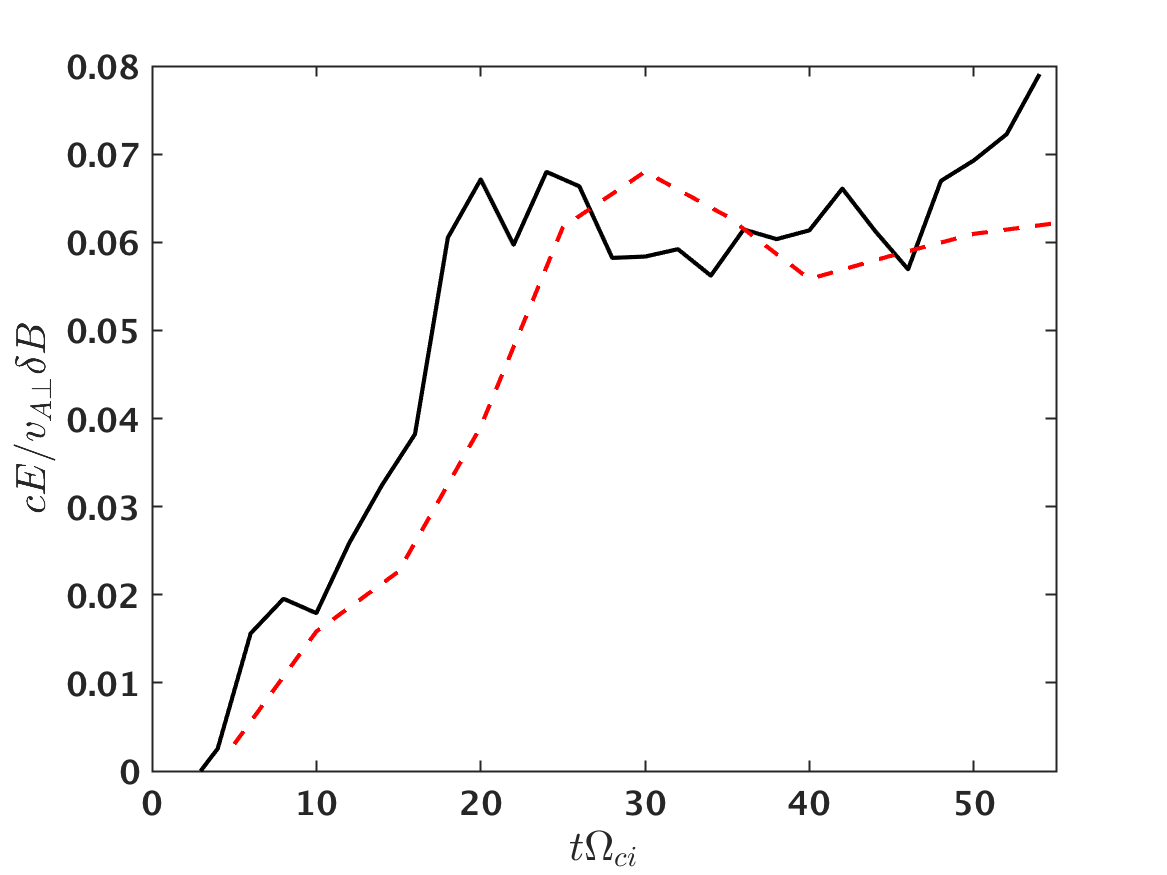}
\includegraphics[width=0.48\linewidth]{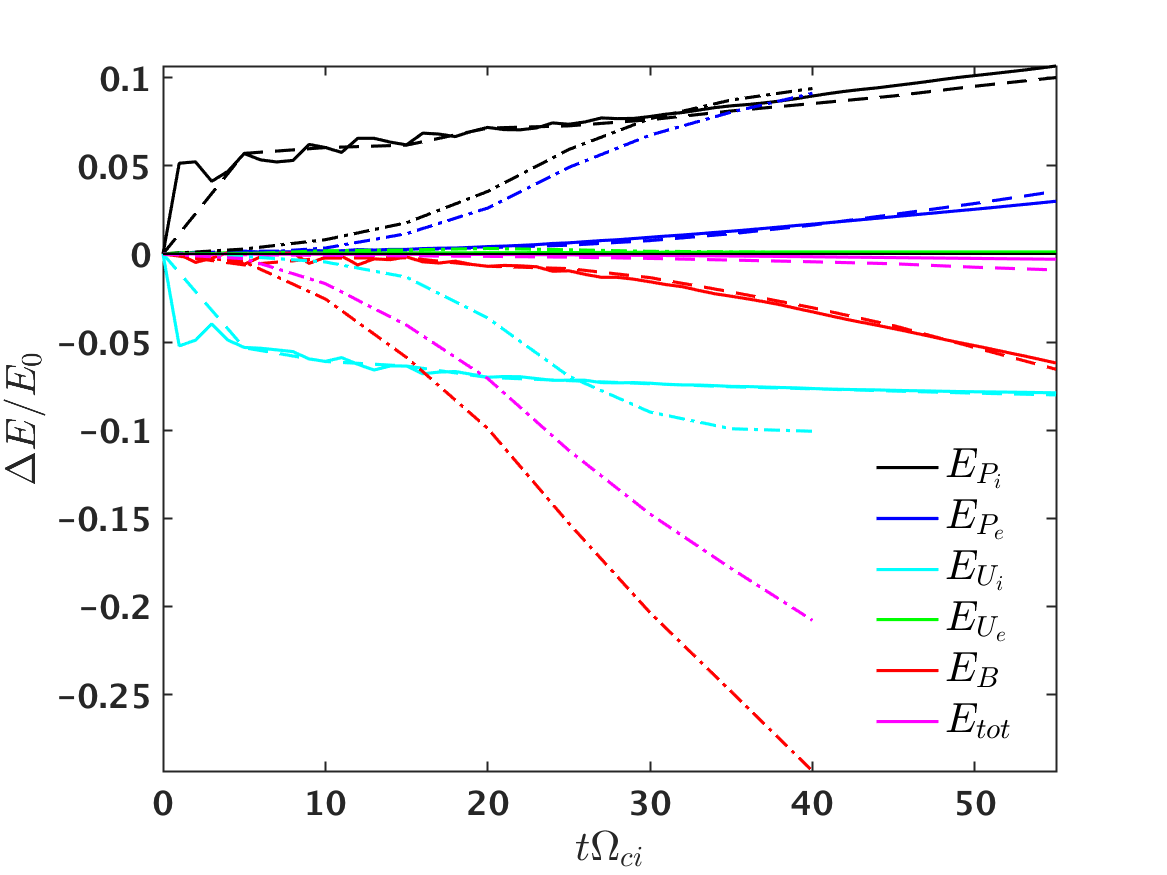}
\caption{(Left) The reconnection rate for runs 2D10 (black) and 3D10 (red dashed). (Right) The relative change in energy for runs 2D10 (solid), 3D10 (dashed), and 3D5 (dash-dotted). $E_0$ is the total initial energy, $E_0 = E_{tot}(t=0)$, and $\Delta E_* = E_*(t) - E_*(t=0)$. }\label{fig:RateEnergy}
\end{figure}

We first present the results of the Burch event study using the ten-moment model in two and three dimensions. In figure~\ref{fig:RateEnergy} is presented the reconnection rates (left panel) and relative changes of energy components (right) for runs 2D10 (solid) and 3D10 (dashed). The reconnection rate in the 3D case was computed by averaging $B_x$ and $B_y$ over the $z$-direction and integrating to construct an average vector potential, $\langle A_z \rangle$. In both 2 and 3D, the rate reaches a plateau at the canonically observed value of $c E / v_{A\perp s} B_{x s} \sim 0.1$. The relative change of the energies is essentially identical in 2D and 3D for the ten-moment model. Thus, in terms of energy exchange and reconnection rates, the 2 and 3D ten-moment simulations are fundamentally similar.

\begin{figure}[t]
\includegraphics[width=0.48\linewidth]{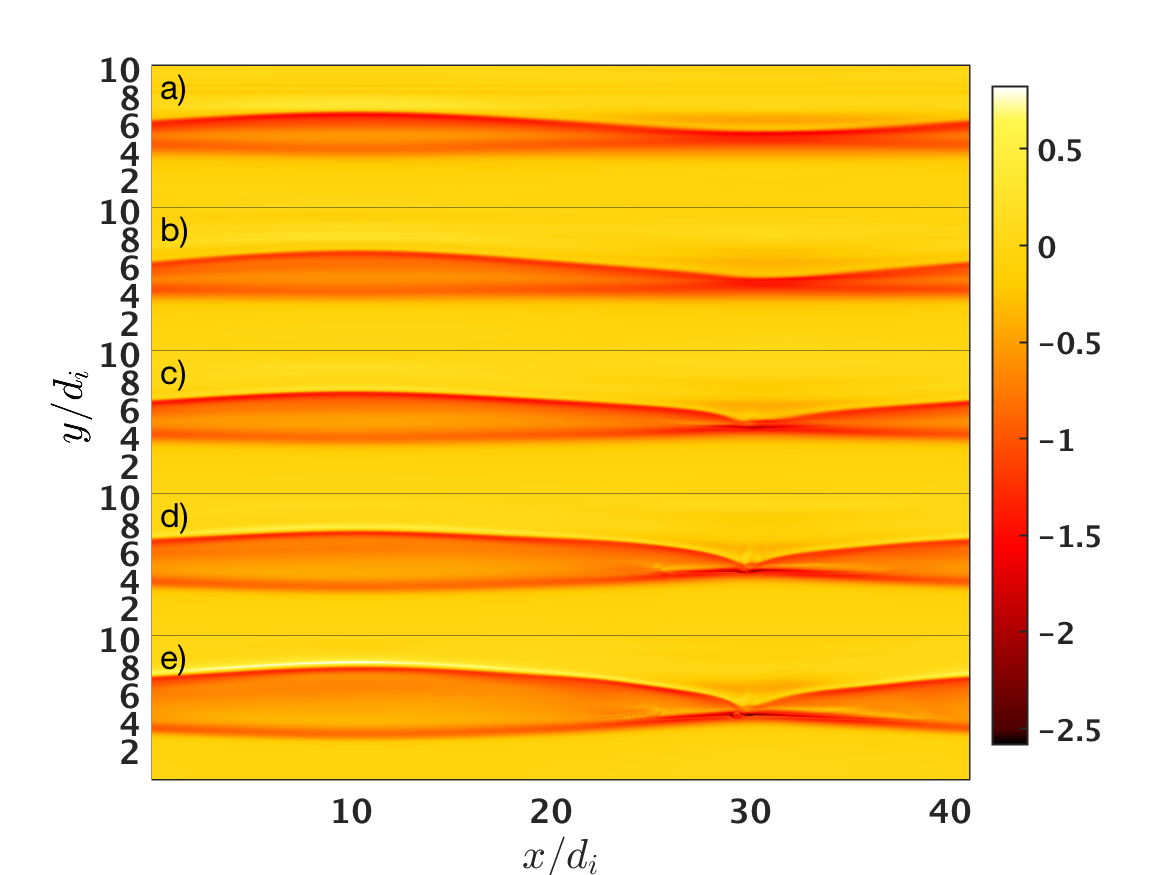}
\includegraphics[width=0.48\linewidth]{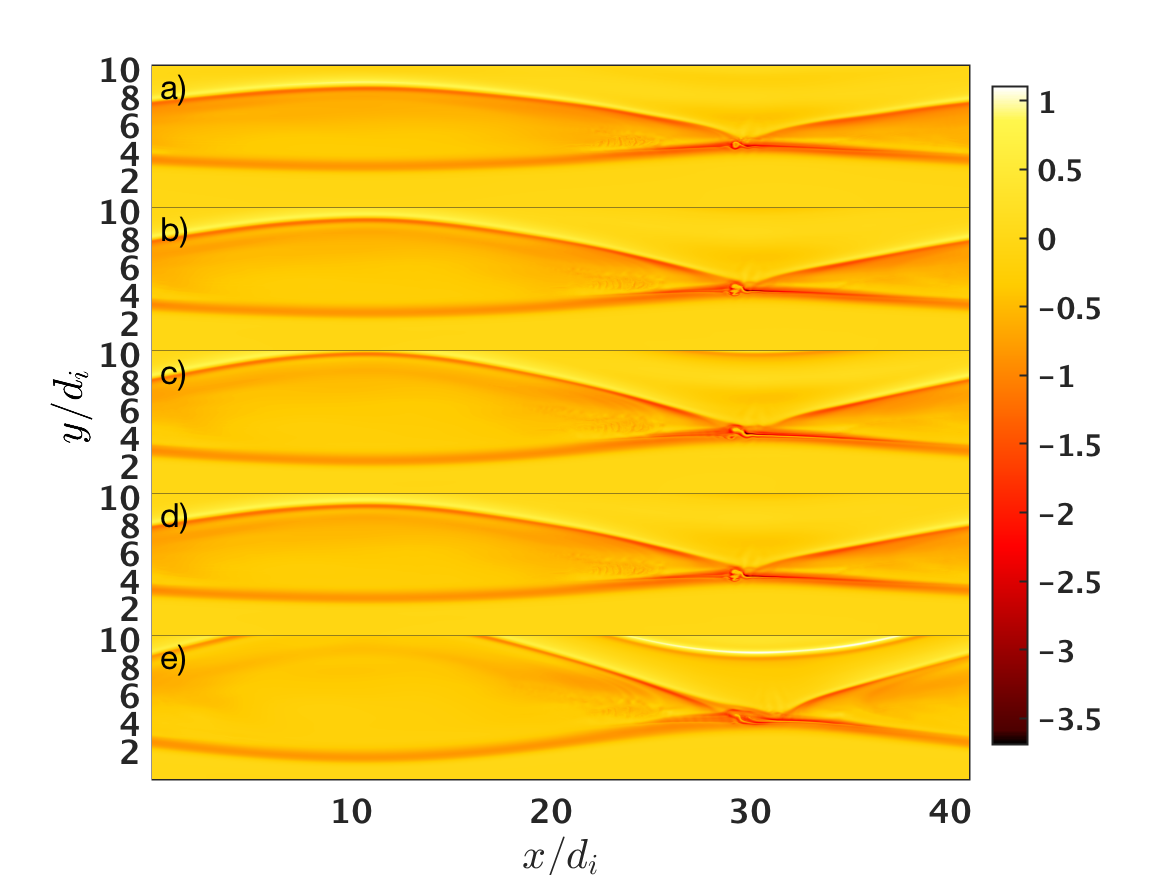}
\caption{Slices of the out-of-plane current density, $J_z$, for run 2D10 at times $t \Omega_{ci} = 10$ to $30$ (left) and $35$ to $55$ (right), with each frame separated by $5/\Omega_{ci}$.}\label{fig:Jz2D10}
\end{figure}

\begin{figure}[t]
\includegraphics[width=0.48\linewidth]{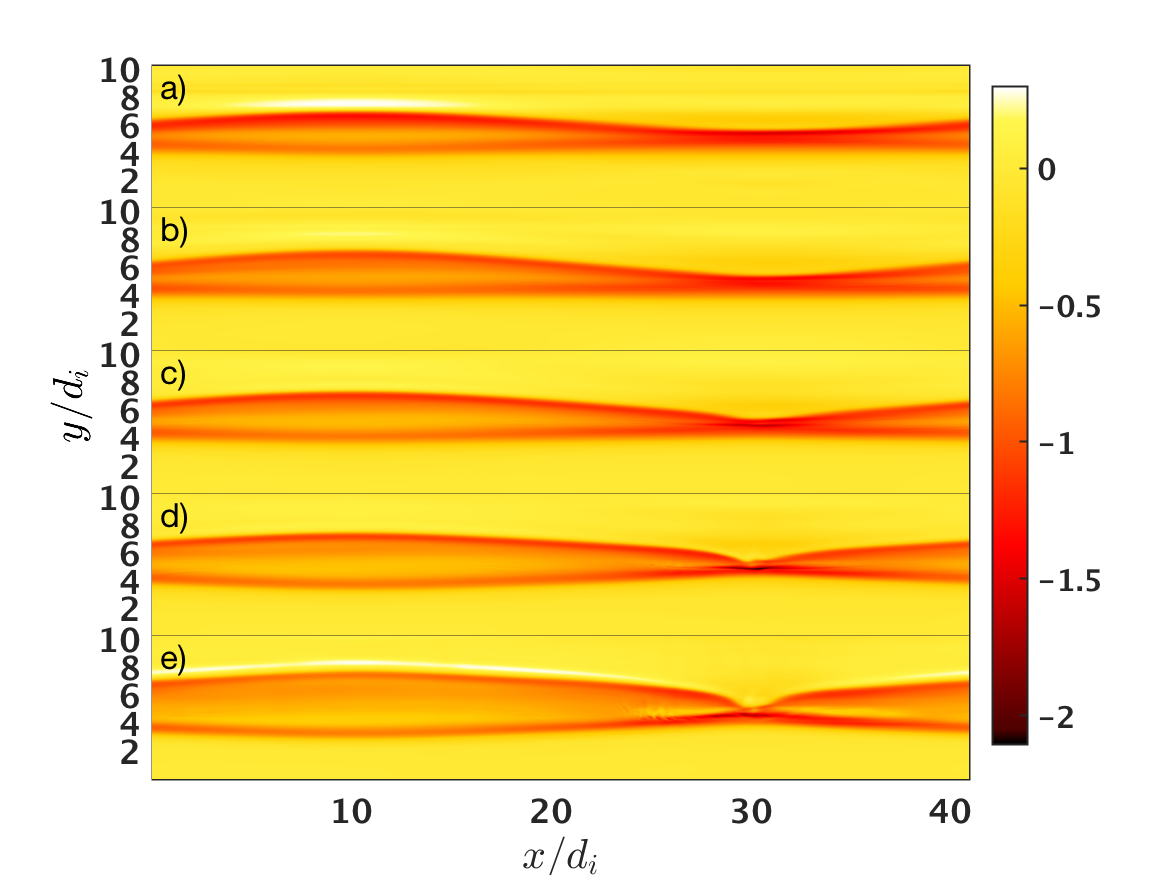}
\includegraphics[width=0.48\linewidth]{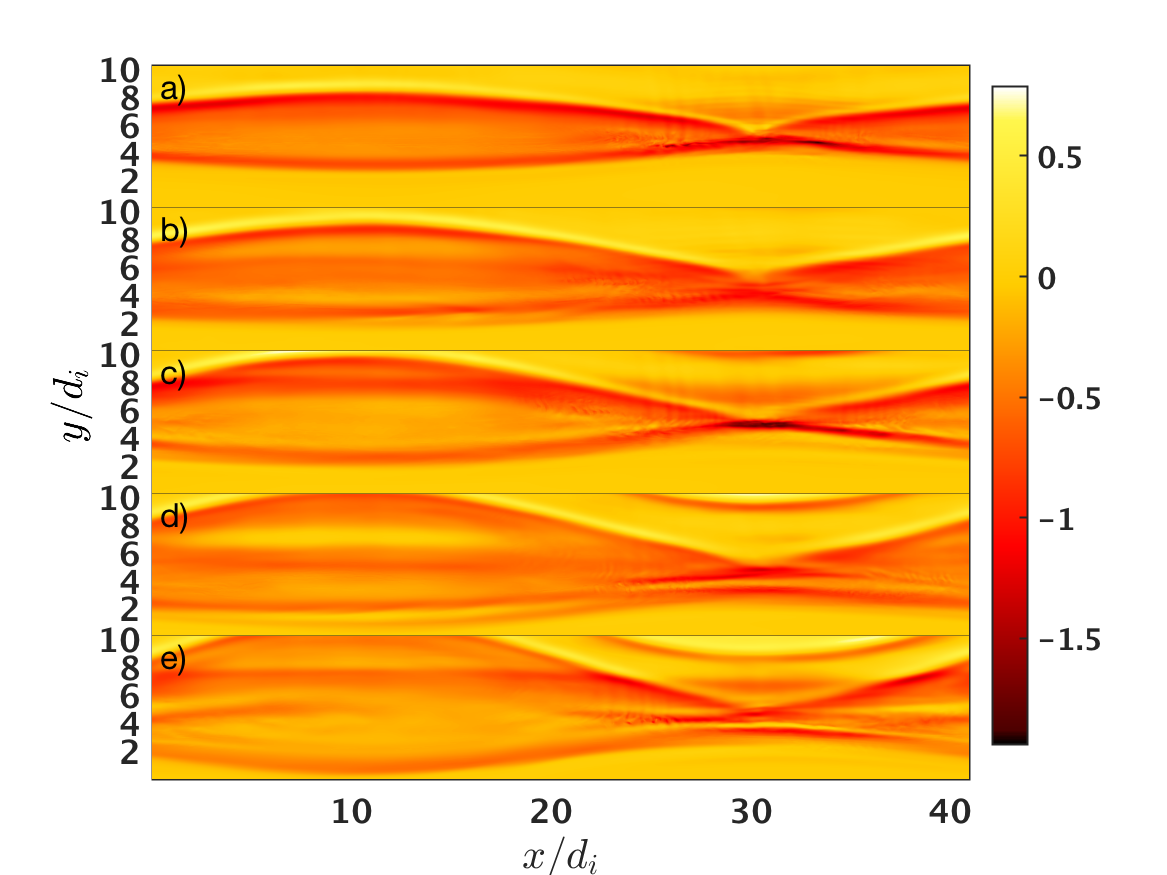}
\caption{Slices of the out-of-plane current density, $J_z$, taken at $z=0$ for run 3D10 at times $t \Omega_{ci} = 10$ to $30$ (left) and $35$ to $55$ (right), with each frame separated by $5/\Omega_{ci}$.}\label{fig:Jz3D10}
\end{figure}

\begin{figure}[t]
\includegraphics[width=\linewidth]{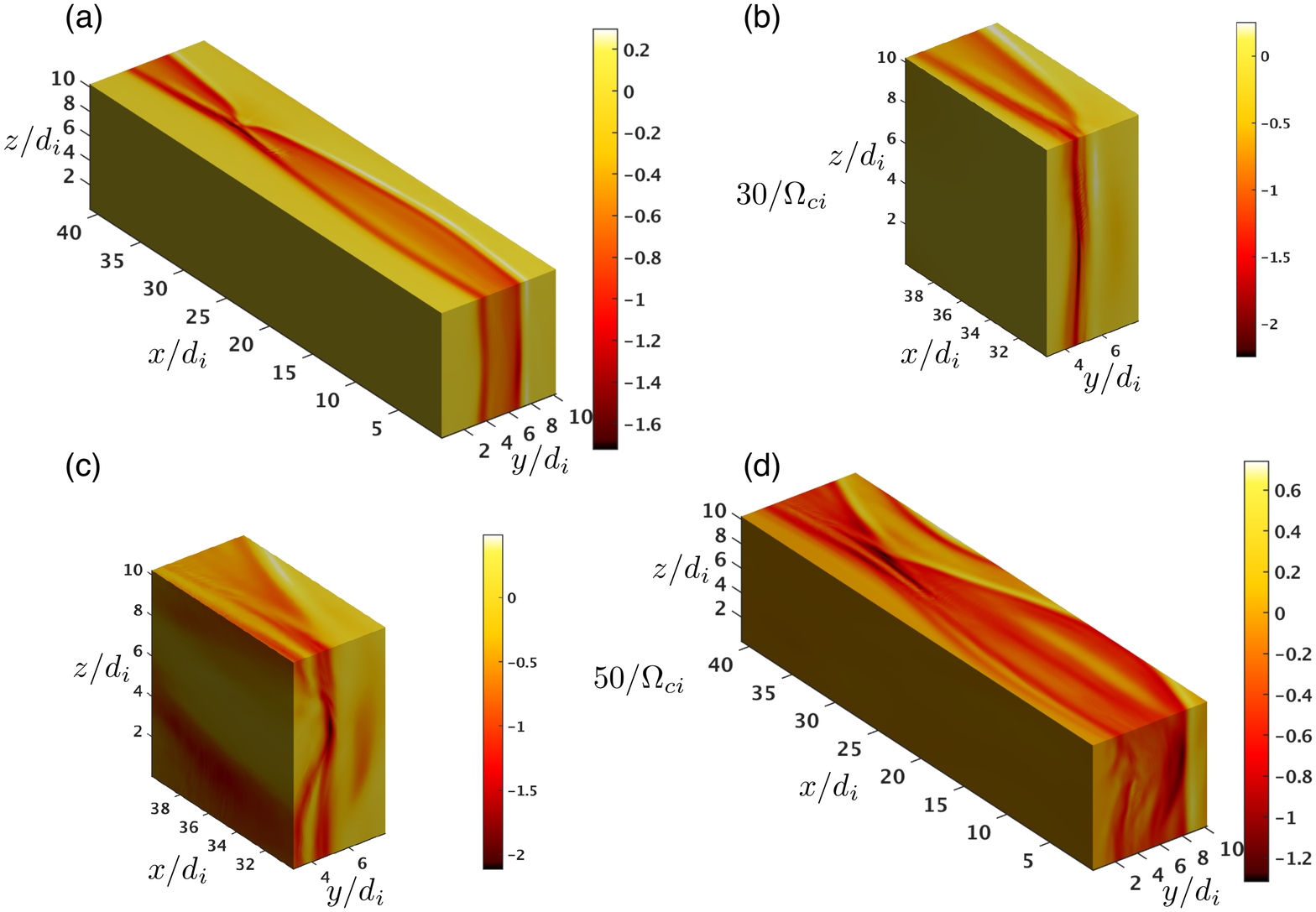}
\caption{The full 3D evolution of the current density $J_z$ in run 3D10. The upper two panels, (a) and (b), correspond to $t \Omega_{ci} = 30$, and the lower two correspond $t \Omega_{ci} = 50$.}\label{fig:10M3D}
\end{figure}

\begin{figure}[t]
\includegraphics[width=0.48\linewidth]{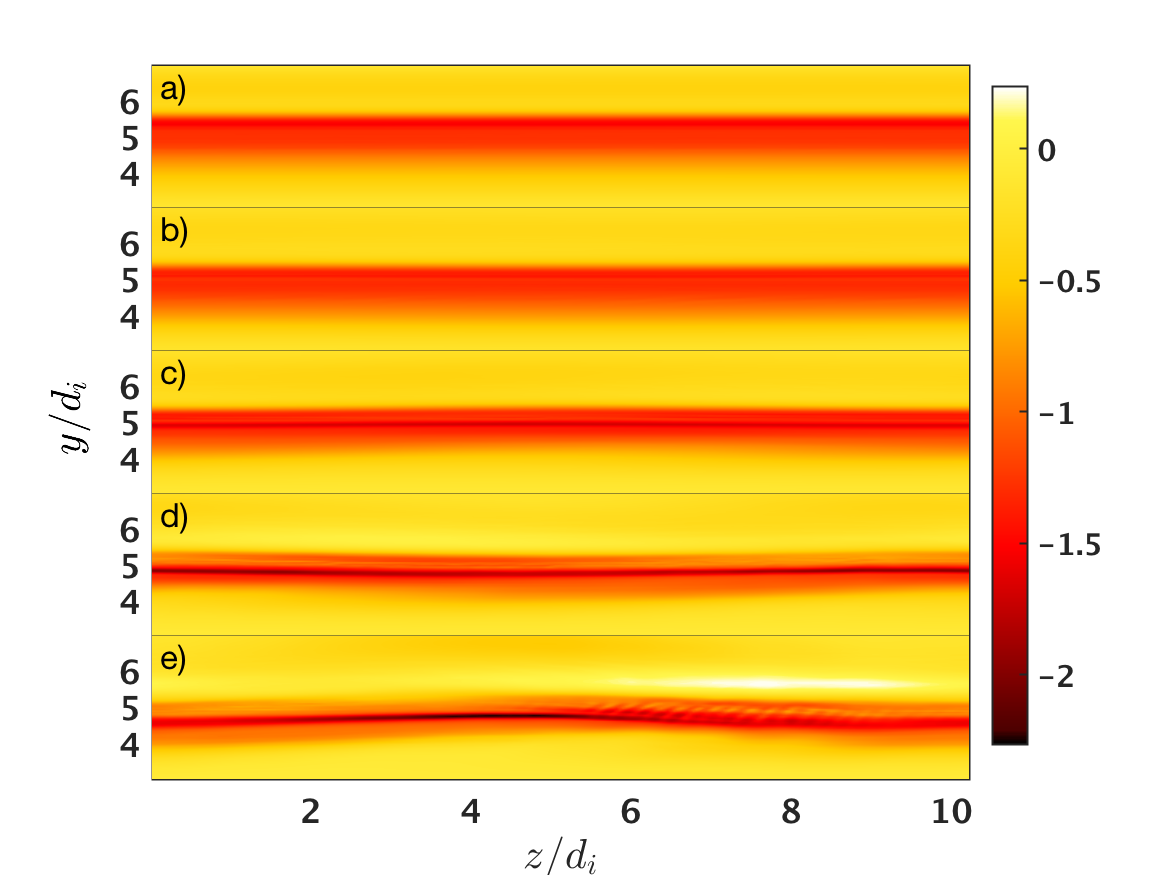}
\includegraphics[width=0.48\linewidth]{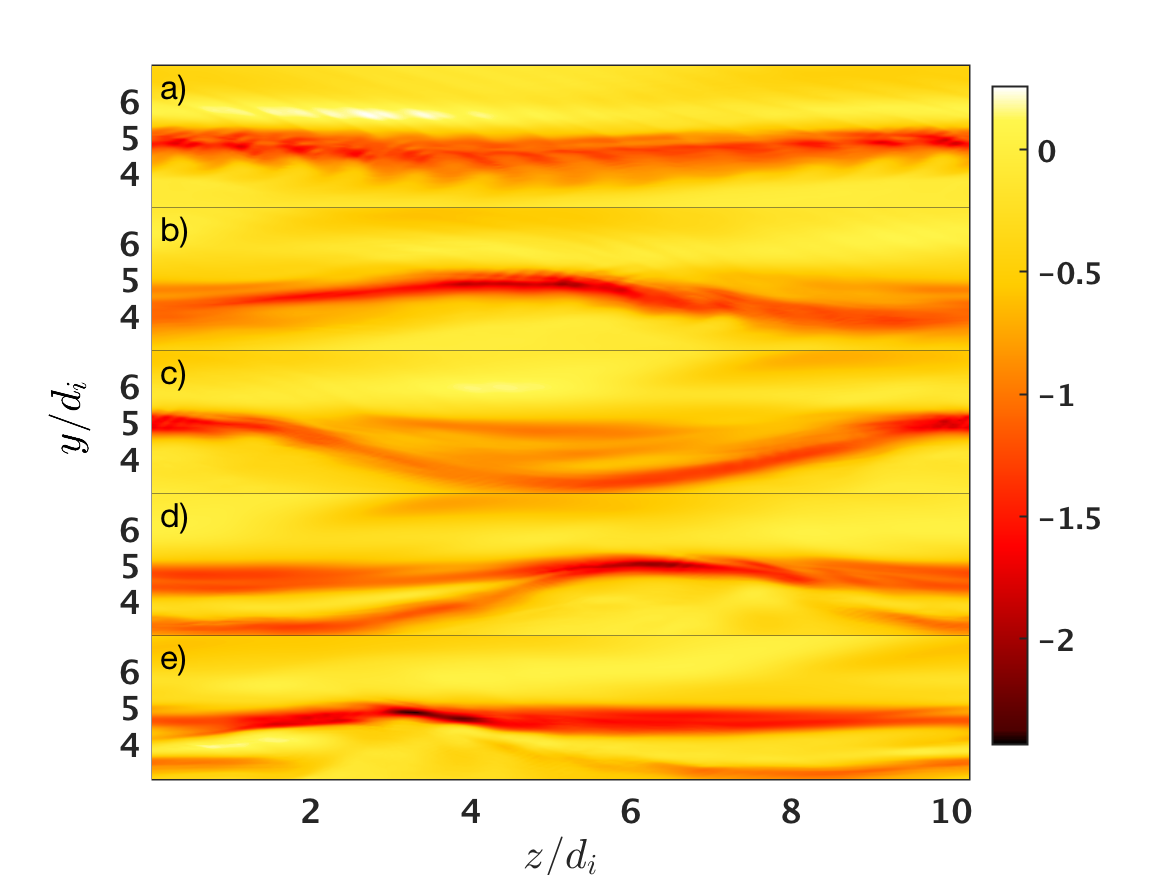}
\caption{Slices of the current density $J_z$ taken at at the X-line, $x=30 d_i$, for run 3D10 at times $t \Omega_{ci} = 10$ to $30$ (left) and $35$ to $55$ (right), with each frame separated by $5/\Omega_{ci}$. }\label{fig:10MJz}
\end{figure}

The evolution of the out-of-plane current density, $J_z$, in the $xy$-plane for runs 2D10 and 3D10 are presented in figures~\ref{fig:Jz2D10} and \ref{fig:Jz3D10}. For run 3D10, the current density is taken at $z=0$. The early-time evolution for $t\Omega_{ci} = 10$ to $30$ in $5/\Omega_{ci}$ increments is presented in the left panel of each figure. Little qualitative or quantitative difference exists during this phase, the most significant difference being a moderately increased current density in 2D due to the existence of sharper magnetic field gradients. The right panels of the figures presents the current density at times $t\Omega_{ci} = 35$ to $55$. Beginning at $t\Omega_{ci} = 35$, the 2 and 3D evolution noticeably diverges. The source of the divergence becomes clear by examining the full 3D evolution of the current in run 3D10, as shown in figure~\ref{fig:10M3D}. At $t\Omega_{ci}=30$ the current layer is predominantly laminar, while at $t\Omega_{ci}=50$, the entire layer has become turbulent. In figure~\ref{fig:10MJz} is shown the current in the $yz$-plane at the X-line, $x=30 d_i$, at the same times as figure~\ref{fig:Jz3D10}. It is clear that in three dimensions turbulence plays an active role, broadening the current layers, thereby further reducing the local current density. The entire current layer is also kink unstable with wavelength equal to the longest wavelength accessible in the domain.

\begin{figure}[t]
\includegraphics[width=0.5\linewidth]{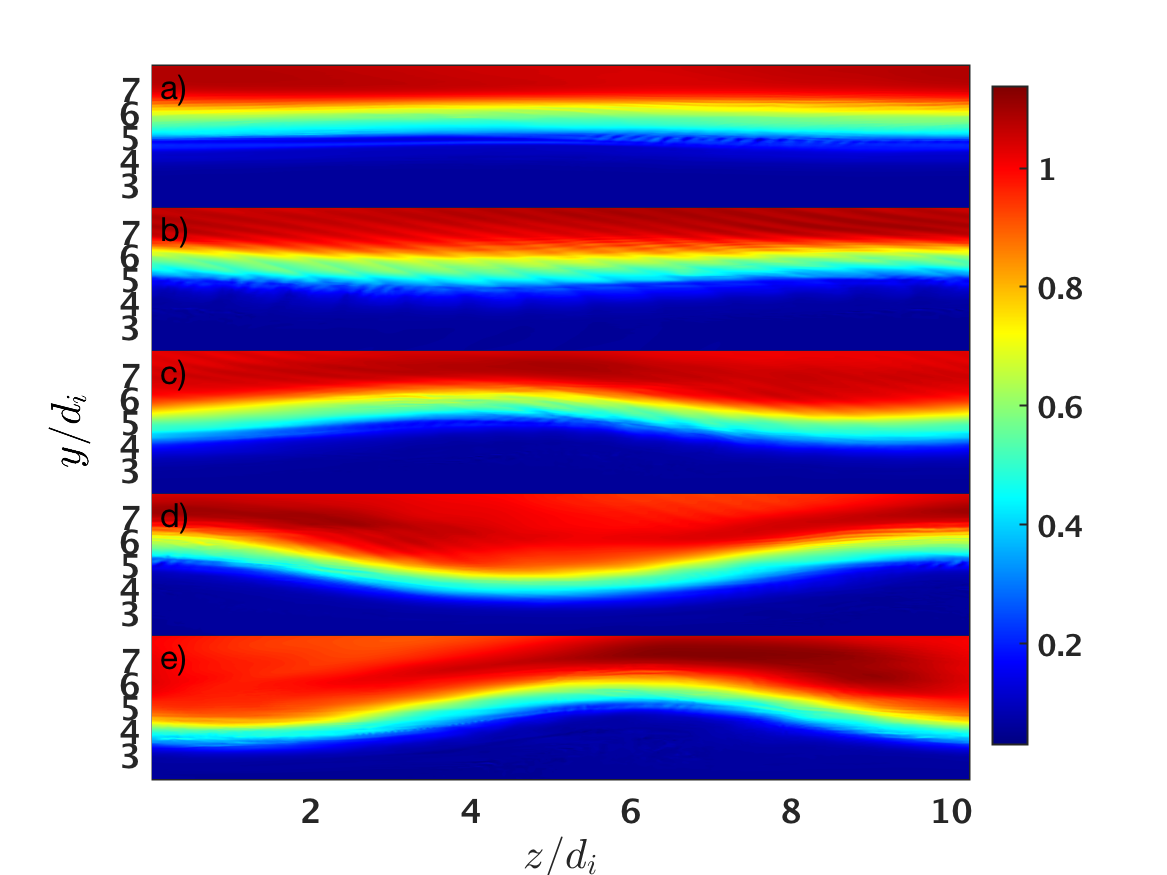}
\includegraphics[width=0.5\linewidth]{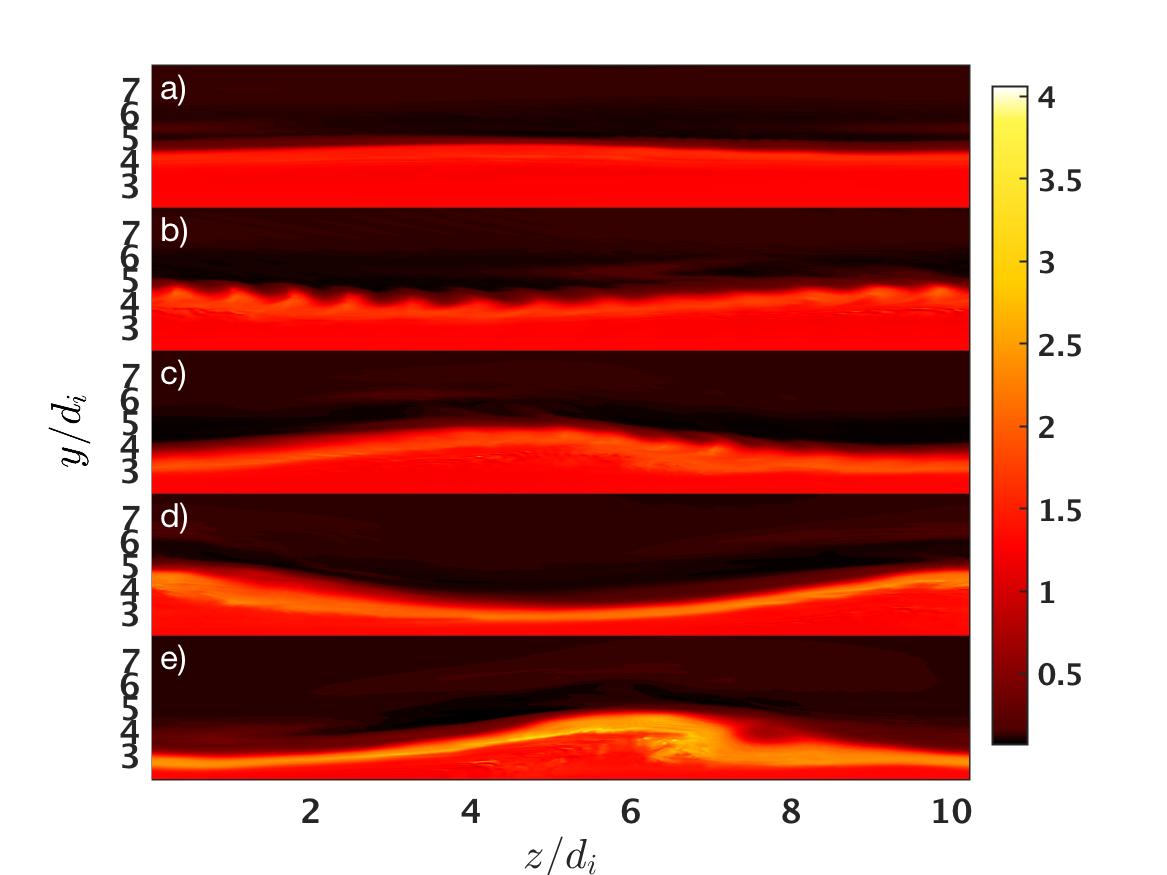}
\caption{Slices of the electron density (left) and electron temperature (right) taken at the X-line, $x=30 d_i$, for run 3D10 at times $t \Omega_{ci} = 30$ to $50$, with each frame separated by $5/\Omega_{ci}$. }\label{fig:10MneTe}
\end{figure}

The turbulence is generated by a field aligned shear flow instability with wavelength $k d_e \simeq 1$ \citep{Romero:1992}. Figure~\ref{fig:10MneTe} presents slices of the electron density (left) and temperature (right) at the X-line at times prior to and following the growth of the instability, $t \Omega_{ci} = 30$ to $50$. The instability drives a low amplitude density perturbation along the current layer that spreads into the high density, magnetosheath, side. By using the large temperature gradient across the layer as a tracer in the right panel of figure~\ref{fig:10MneTe}, it becomes apparent that the instability drives Kelvin-Helmholtz like vorticies along the current layer that lead to enhanced mixing relative to the 2D simulation. This instability is active along the entire separatrix layer; however, it is most significant in the vicinity of the X-line. 

The electron shear flow instability present in run 3D10 has been observed in PIC simulations of the Burch event \citep{Le:2017}; however, the shear instability is preceded by the growth of the lower hybrid drift instability (LHDI) \citep{Krall:1971,Daughton:2003}, which feeds off of the large density gradient present in this event, but the LHDI is absent from this ten-moment simulation. Prior particle-in-cell simulations of the Burch event \citep{Price:2016,Le:2017} find that the LHDI plays a fundamental role in generating turbulence that both supports the reconnection electric field and enhances mixing and electron heating. Linear analysis of the ten-moment model performed in \citet{Ng:2019} shows that the LHDI is indeed captured by the model, but the closure parameter for both species, $k_{0s}$, can significantly modify the growth rate of the instability. To recover kinetic growth rates, the \textit{ion} closure parameter must be approximately equal to the wavelength of the fastest growing LHDI mode. Unfortunately, the fastest growing LHDI modes have $k \rho_e \sim 1$ \citep{Daughton:2003}, which differs significantly from the closure parameter that is necessary to model reconnection in the vicinity of the X-line, $k_{0i} d_i \lesssim 1$, and the value chosen for this study. Therefore, the ten-moment closure for the parameters employed in run 3D10 damps the LHDI, leading to a growth rate that is reduced by approximately an order of magnitude relative to kinetic theory.

\begin{figure}[t]
\includegraphics[width=0.48\linewidth]{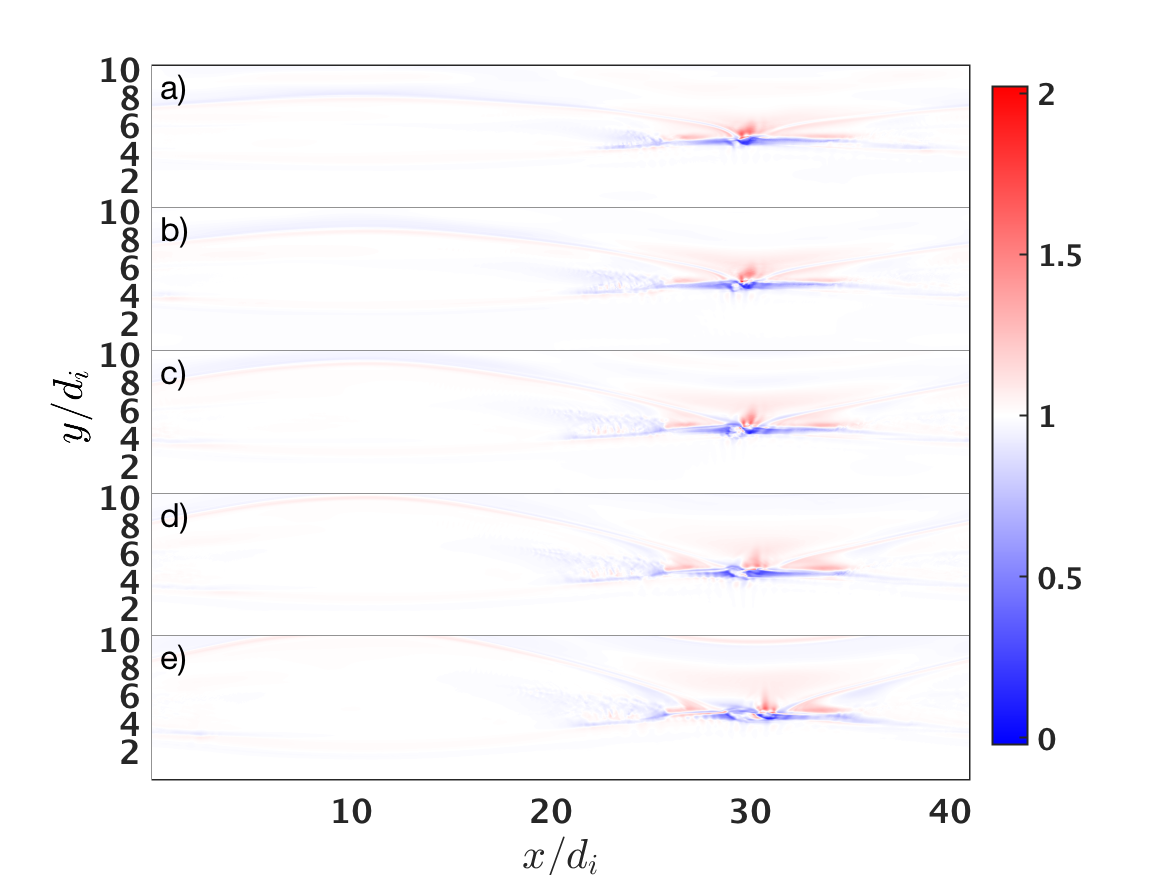}
\includegraphics[width=0.48\linewidth]{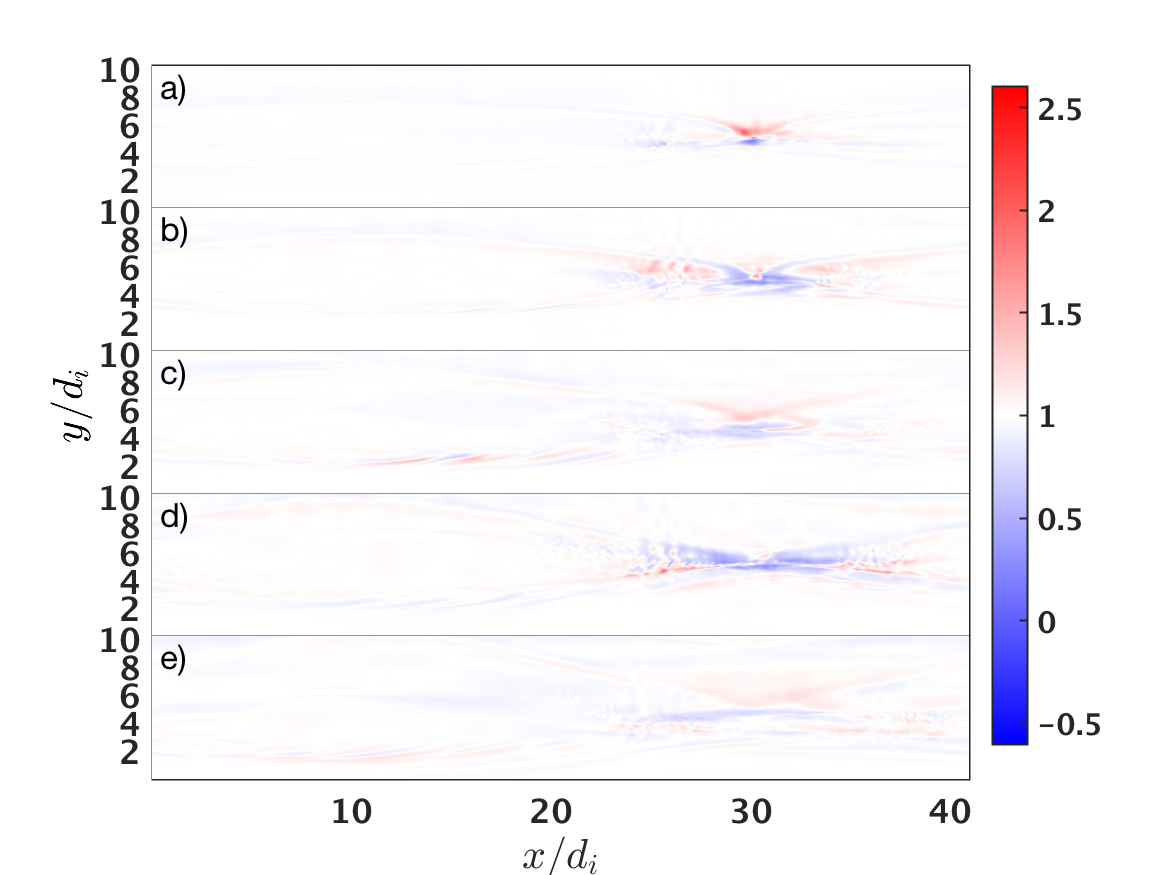}
\caption{Slices of the electron temperature anisotropy, $T_\parallel/T_\perp$ for run 2D10 (left) and 3D10 (right), taken at $z=0$, times $t \Omega_{ci} = 30$ to $50$, with each frame separated by $5/\Omega_{ci}$. }\label{fig:10Aniso}
\end{figure}

\begin{figure}[t]
\includegraphics[width=0.48\linewidth]{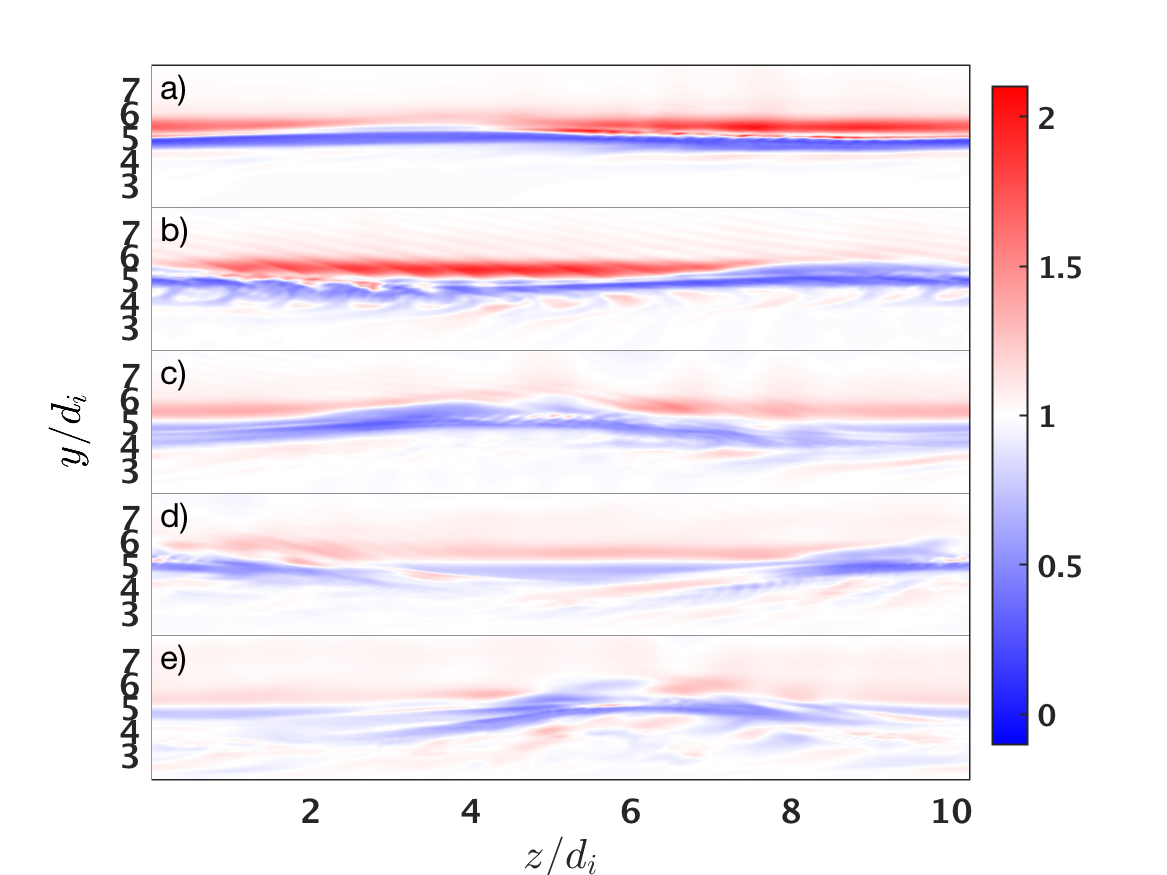}
\includegraphics[width=0.48\linewidth]{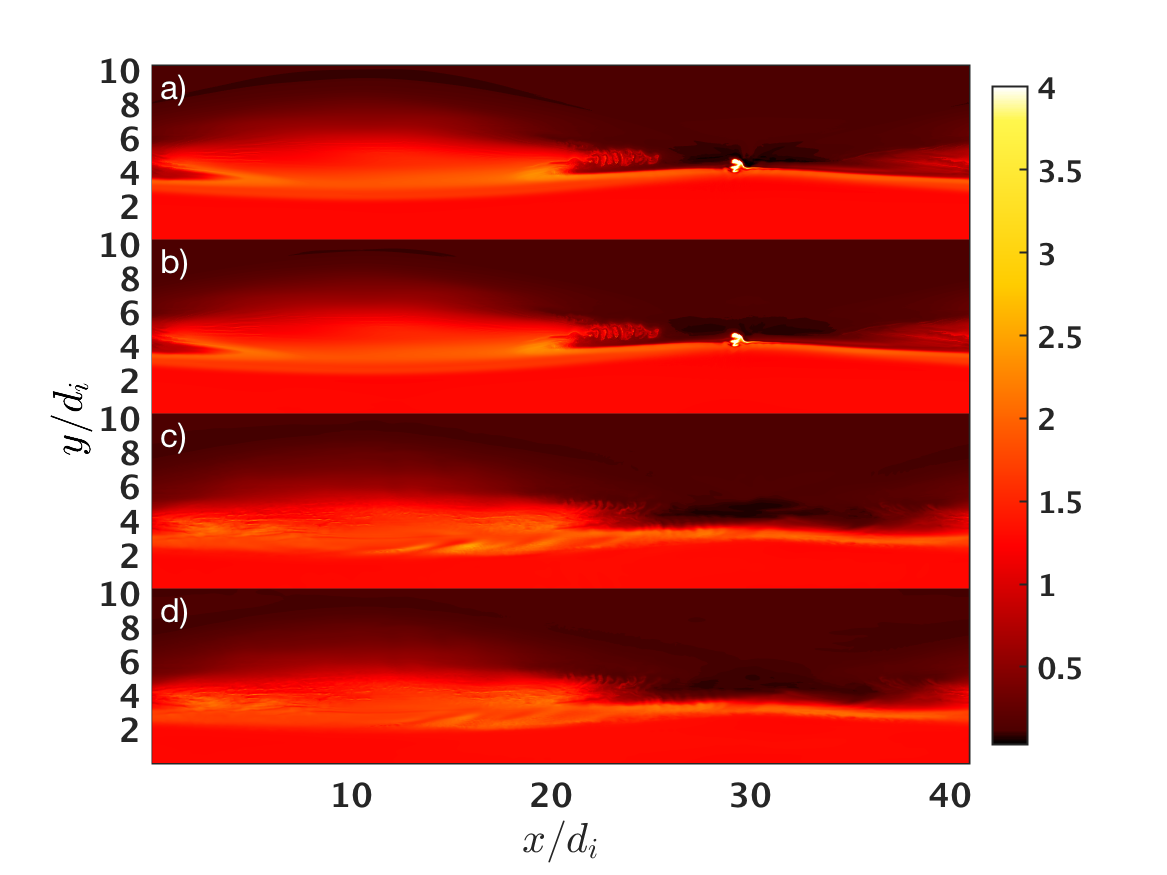}
\caption{(Left) Slices of the electron temperature anisotropy, $T_\parallel/T_\perp$, for run 3D10 taken at the X-line, $y=30 d_i$ at times $t \Omega_{ci} = 30$ to $50$, with each frame separated by $5/\Omega_{ci}$. (Right) Slices of the the parallel (a \& c) and perpendicular (b \& d) electron temperatures for runs 2D10 (a \& b) and 3D10 (c \& d) at time $t \Omega_{ci} = 40$ and $z=0$ for run 3D10.}\label{fig:10AnisoComp}
\end{figure}

The electron temperature anisotropy, $T_\parallel/T_\perp$, for runs 2D10 (left) and 3D10 (right) are presented in figure~\ref{fig:10Aniso} for times  $t \Omega_{ci} = 30$ to $50$. Significant electron temperature anisotropy develops in both two and three dimensions in the ten-moment model; however, the presence of shear driven turbulence and mixing in three dimensions enhances the temperature anisotropy along the separatrices and significantly broadens the region with $T_\perp > T_\parallel$ in the vicinity of the X-line. Examining the structure of the temperature anisotropy in the $yz$-plane in the left panel of figure~\ref{fig:10AnisoComp} for run 3D10 reveals that the most significant excess parallel heating occurs in regions where the current layer has oscillated toward the magnetosphere, creating local depressions of the magnetic field amplitude. 

The largest value of the temperature anisotropy occurs in run 3D10 and is $T_\parallel/T_\perp \simeq 2.5$, which is in close agreement with the value observed by MMS [see figure 3I of \citet{Burch:2016}]. However, the Burch event observation of $T_\parallel/T_\perp > 1$ is entirely on the low density, magnetosphere side of the current layer, and the magnetosheath is observed to be nearly isotropic. In both runs 3D10 and and 2D10, $T_\parallel/T_\perp > 1$ is only generated on the magnetosheath side of the X-line, while the magnetosphere side is predominantly $T_\parallel/T_\perp < 1$ plasma. The right panel of figure~\ref{fig:10AnisoComp} presents electron $T_\parallel$ (a \& c) and $T_\perp$ (b \& d) separately for runs 2D10 (a \& b) and 3D10 (c \& d) at $t\Omega_{ci} = 40$ and $z=0$. Overall, the electron heating is isotropic, with regions of anisotropic heating being due to isolated and small variations in the temperature components. Three dimensional PIC simulations of the Burch event find that the entire magnetosphere separatrix layer is heated in only the parallel direction with $T_\parallel/T_\perp \sim 4-5$ [see figure 3 of \citet{Le:2017}], which is consistent with the location but not magnitude of observed temperature anisotropy in the MMS event. The anisotropic heating in the 3D PIC result is attributed to mixing induced by the LHDI, which is only present in three dimensions; however, as noted above, the LHDI is damped in the ten-moment model.

\begin{figure}[t]
\includegraphics[width=0.48\linewidth]{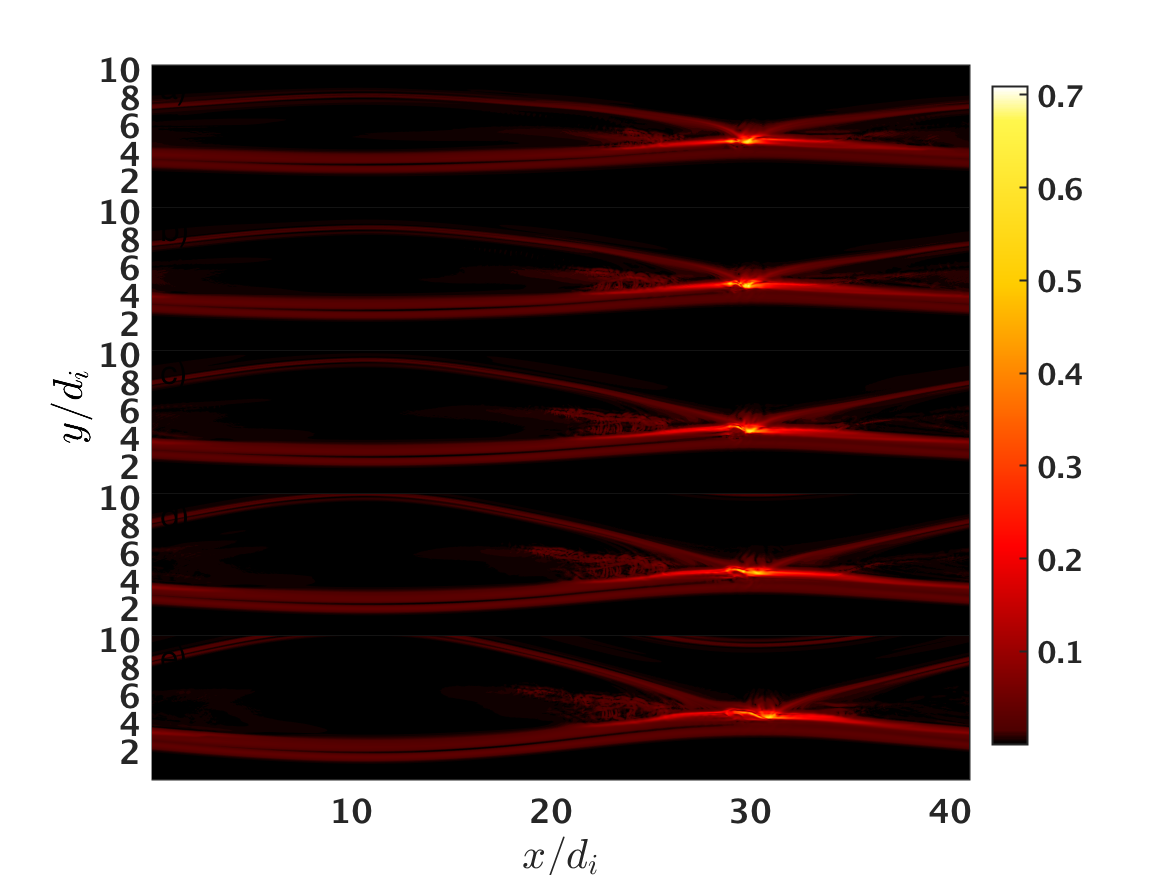}
\includegraphics[width=0.48\linewidth]{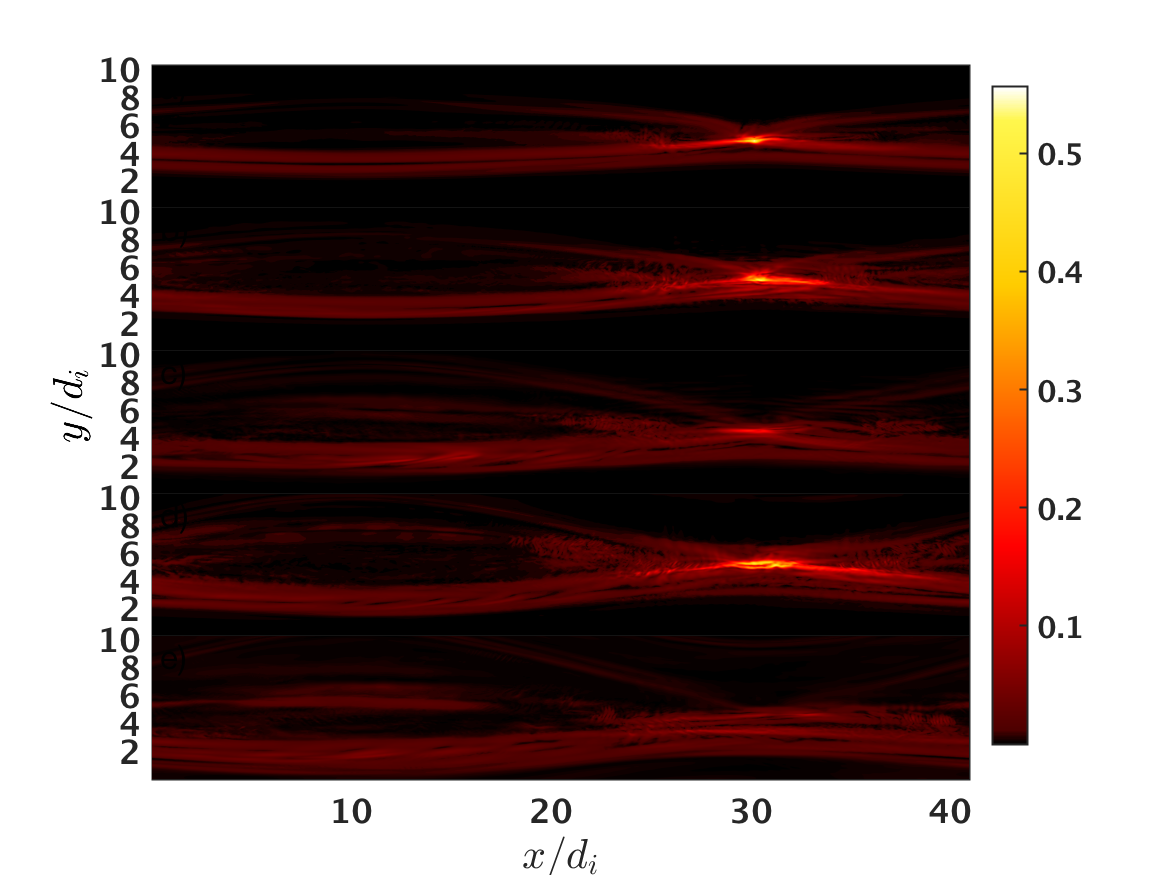}
\caption{Slices of the electron pressure agyrotropy, $\sqrt{Q}$ for run 2D10 (left) and 3D10 (right), taken at $z=0$, times $t \Omega_{ci} = 30$ to $50$, with each frame separated by $5/\Omega_{ci}$. }\label{fig:10MAgyro}
\end{figure}

Since the ten-moment model evolves the full pressure tensor, we present in figure~\ref{fig:10MAgyro} the electron pressure agyrotropy $\sqrt{Q}$ for runs 2D10 (left) and 3D10 (right), where
\begin{equation}
    Q = \frac{P_{12}^2 + P_{13}^2 + P_{23}^2}{P_\perp^2 + 2P_\perp P_\parallel} 
\end{equation}
is the measure of gyrotropy defined in \citet{Swisdak:2016}. By definition, $0 \le Q \le 1$, where $Q = 0 (1)$ corresponds to gyrotropy (maximal agyrotropy). The peak values of $\sqrt{Q} \simeq 0.7$ and $0.55$ occur in the vicinity of the X-point and X-Line in 2 and 3D, respectively, with the reduced peak value in run 3D10 due to turbulence broadening the current layers. The gyrotropy measure also highlights the separatricies, where $\sqrt{Q} \simeq 0.05$. The structure and magnitude of $\sqrt{Q}$ are generally consistent with results observed in PIC simulations and \textit{in situ} spacecraft observations of reconnection, e.g., \citet{Swisdak:2016,Zenitani:2016,Bourdin:2017,Genestreti:2018}.

\begin{figure}[t]
\includegraphics[width=0.48\linewidth]{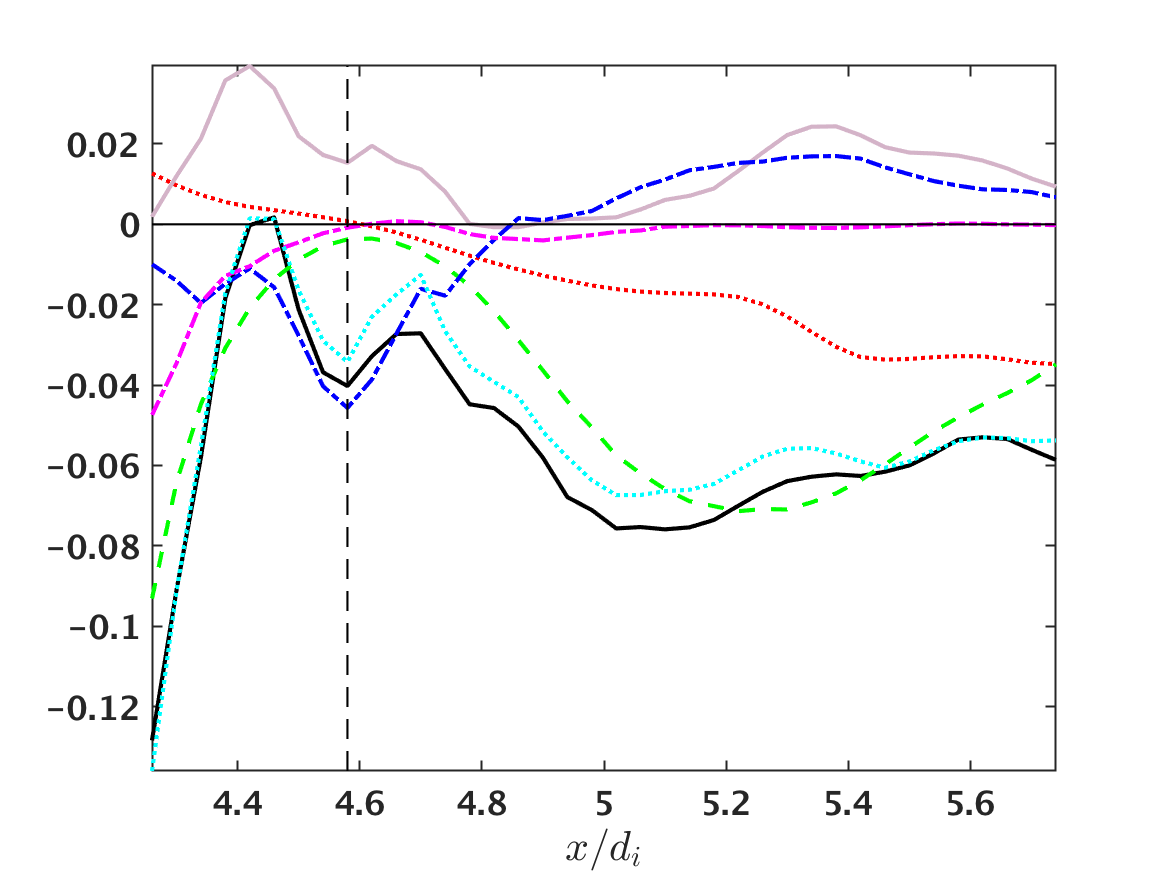}
\includegraphics[width=0.48\linewidth]{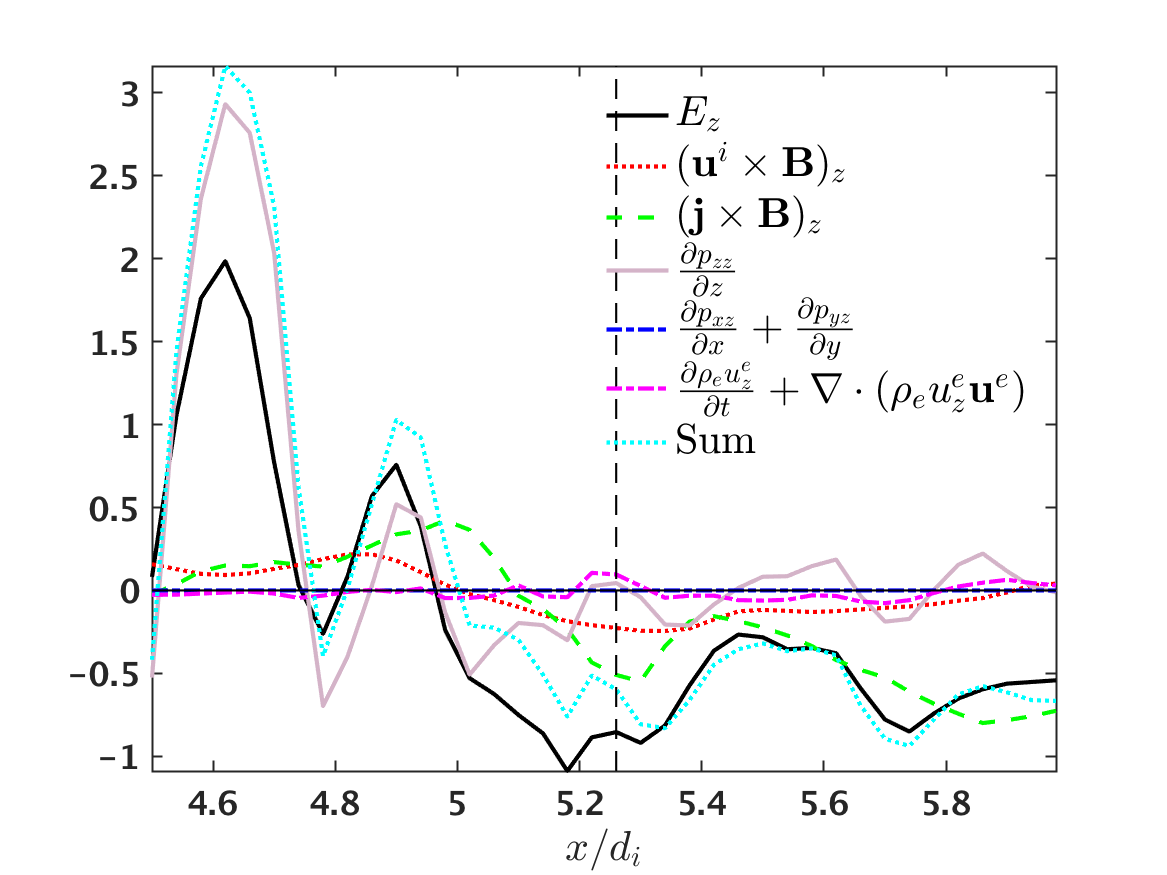}
\caption{The 1D components of the generalized Ohm's law across the current layer for runs 3D10 at time $t \Omega_{ci} = 40$ (left) and 3D5 at $t \Omega_{ci} = 20$ (right), both slices taken at $z=0$. The vertical dashed line in each figure corresponds to location of the magnetic field reversal.}\label{fig:OhmsLaw}
\end{figure}

To further emphasize the importance of the pressure agyrotropy at the X-line, we next examine the components of the generalized Ohm's law for the ten-moment model, which can be derived from the electron momentum equation, equation~\eqref{eq:u}. At the X-line, only the guide field, $z$, aligned component of the Ohm's law contributes
\begin{equation}\label{eq:OhmsLaw}
\begin{split}
    E_z = &-(\V{u}^i \times \V{B})_z + \frac{1}{n_e |e|}(\V{j}\times \V{B})_z - \frac{1}{n_e |e|}\left(\frac{\partial P_{xz}^e}{\partial x} + \frac{\partial P_{yz}^e}{\partial y} + \frac{\partial P_{zz}^e}{\partial z}\right)\\
    &- \frac{1}{n_e |e|}\left(\frac{\partial \rho_e u_z^e}{\partial t} + \nabla \cdot (\rho_e u_z^e \V{u}^e) \right),
\end{split}
\end{equation}
where $e$ is the charge of an electron and $\rho_e = m_e n_e$ is the electron mass density. The one dimensional components of the generalized Ohm's law across the current layer for run 3D10 at time $t \Omega_{ci} = 40$ and $z=0$ are plotted in the left panel of figure~\ref{fig:OhmsLaw}, where the vertical dashed line indicates the location of the magnetic field reversal. The only component of the Ohm's law that contributes significantly at the reversal is the off-diagonal pressure tensor (dot dashed blue). Away from the X-line, the Hall term, $\V{j} \times \V{B}$, (dashed green) is the primary contributor.

\subsection{Five-moment Simulation Results}\label{sec:resultsFive}

Moving to the five-moment model, we begin by examining the relative change in energy for run 3D5 presented in the right panel of figure~\ref{fig:RateEnergy} as dash-dotted lines. The evolution of run 3D5 is notably different from the ten-moment simulations, with the magnetic field decaying rapidly, and its free energy going to ion and electron thermal energy. Also notable is the poor conservation of total energy (magenta) in this run that onsets shortly following the initial decay of magnetic energy. Poor conservation of energy indicates that a significant amount of energy is reaching the grid scale, even at this early stage of the simulation. The reconnection rate for run 3D5 is not plotted in this figure for reasons that will become apparent below.

\begin{figure}[t]
\includegraphics[width=0.48\linewidth]{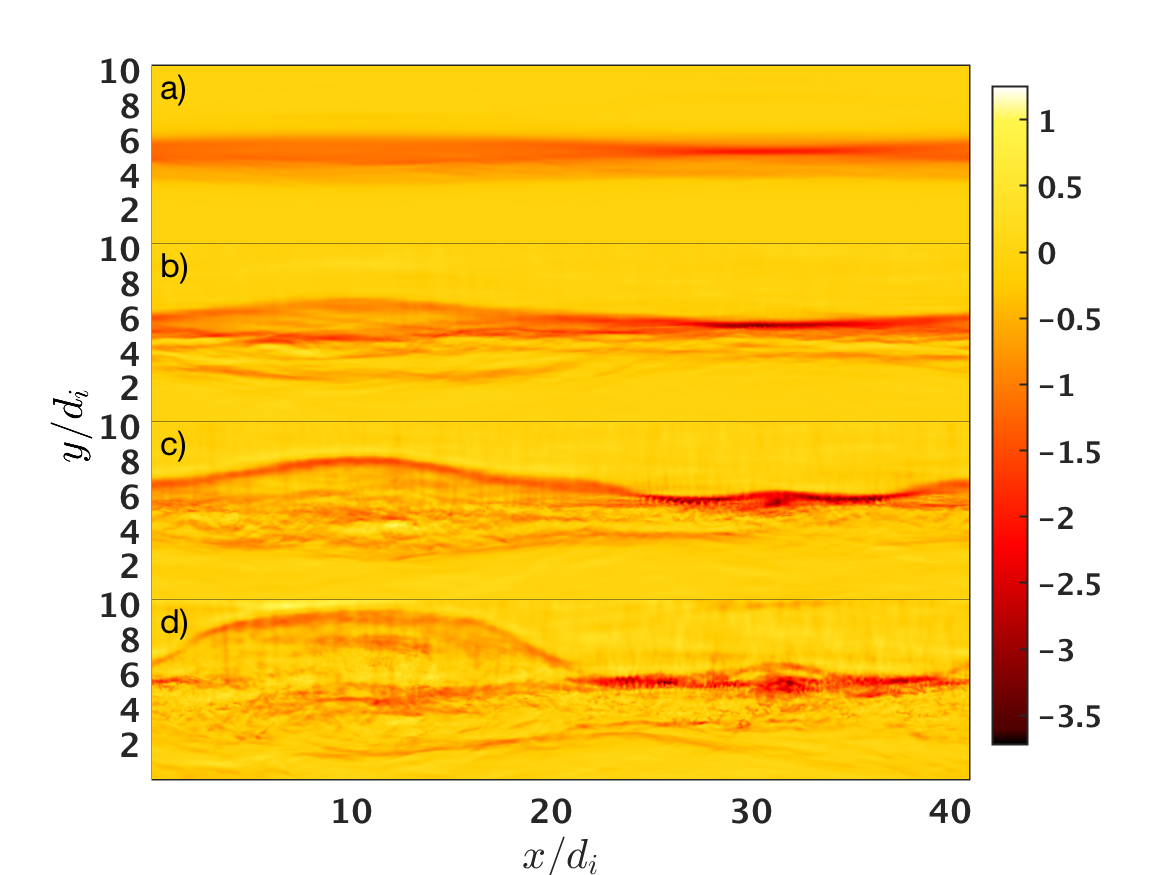}
\includegraphics[width=0.48\linewidth]{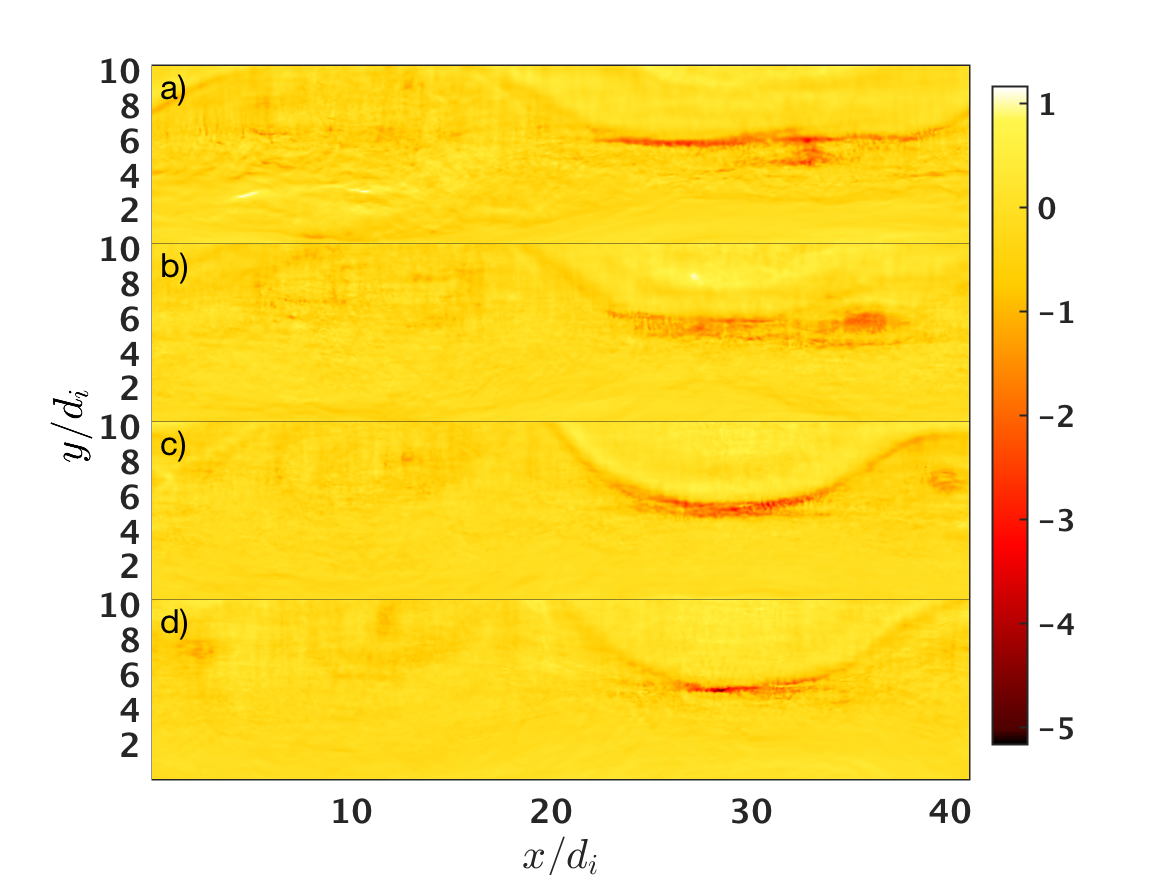}
\caption{Slices of the out-of-plane current density, $J_z$, taken at $z=0$ for run 3D5 at times $t \Omega_{ci} = 5$ to $20$ (left) and $25$ to $40$ (right), with each frame separated by $5/\Omega_{ci}$.}\label{fig:Jz3D5}
\end{figure}

\begin{figure}[t]
\includegraphics[width=\linewidth]{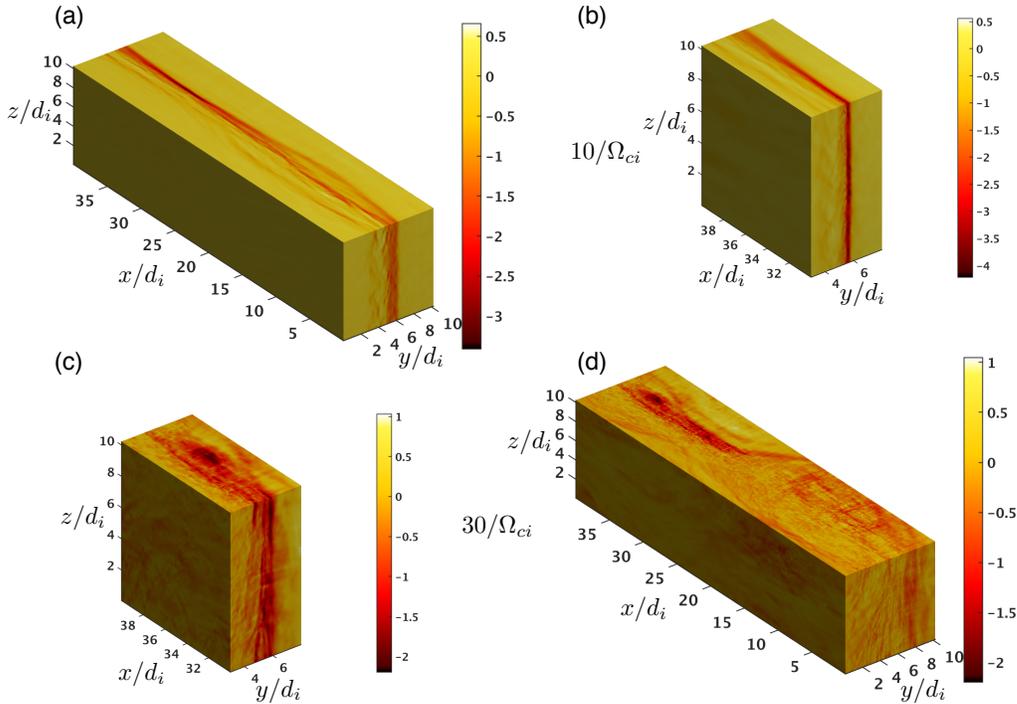}
\caption{The full 3D evolution of the current density $J_z$ in run 3D5. The upper two panels, (a) and (b), correspond to $t \Omega_{ci} = 10$, and the lower two correspond $t \Omega_{ci} = 30$.}\label{fig:5M3D}
\end{figure}

In figure~\ref{fig:Jz3D5} is plotted the evolution of the out-of-plane current density taken at $z=0$ for run 3D5 at times $t \Omega_{ci} = 5$ to $20$ (left) and $25$ to $40$ (right), with each frame separated by $5/\Omega_{ci}$. The full three dimensional structure of the layer at times $t \Omega_{ci} = 10$ and $30$ are presented in figure~\ref{fig:5M3D}. Clearly, the behavior of the five-moment model for these highly asymmetric parameters differs significantly from that of the ten-moment model. The current layer is permeated by turbulence, which leads to substantial diffusion and broadening across the entire layer, essentially destroying the current layer. By $t\Omega_{ci} = 20$, the entire domain has become turbulent. The strong turbulence present in this simulation is responsible for the transport of energy to the grid scale that results in poor energy conservation. Additionally, the turbulence precludes one from reliably calculating the reconnection rate using conventional methods, since the magnetic X and O points are ambiguous. 

\begin{figure}[t]
\includegraphics[width=0.48\linewidth]{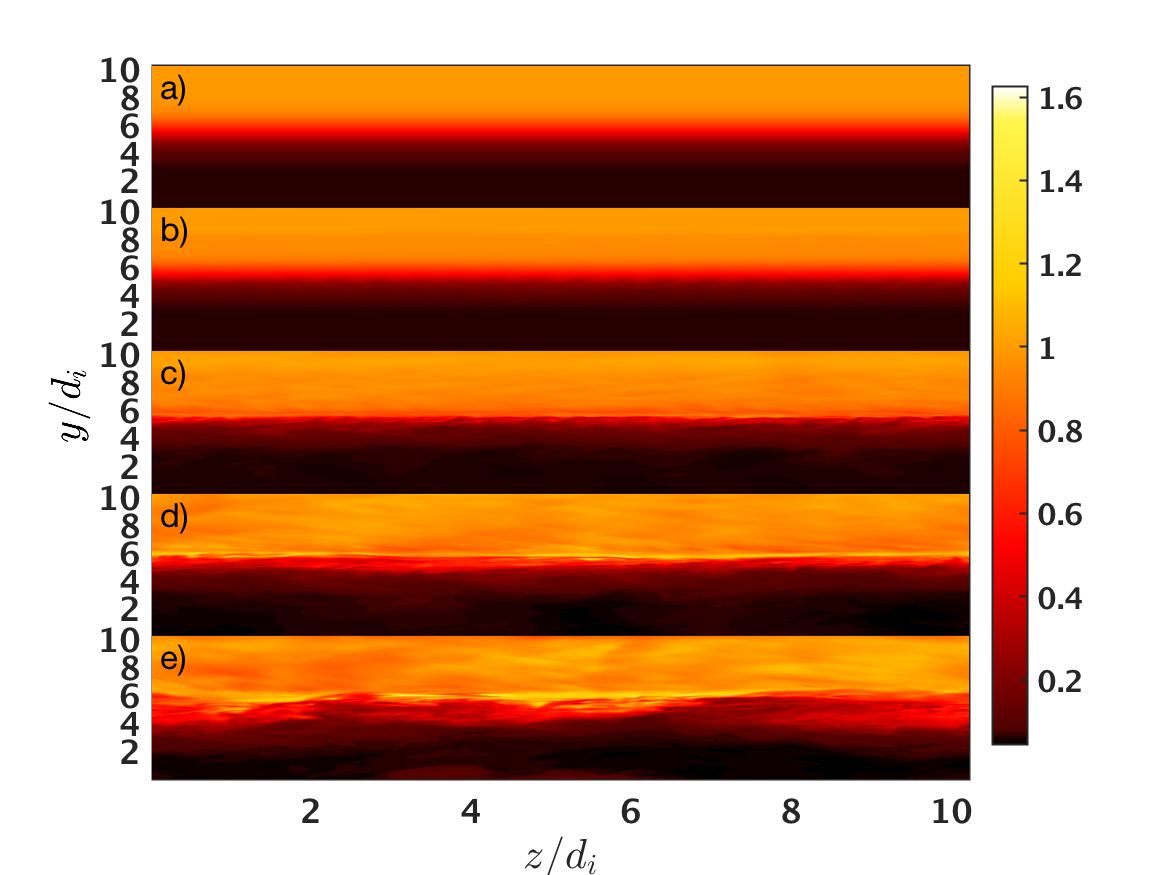}
\includegraphics[width=0.48\linewidth]{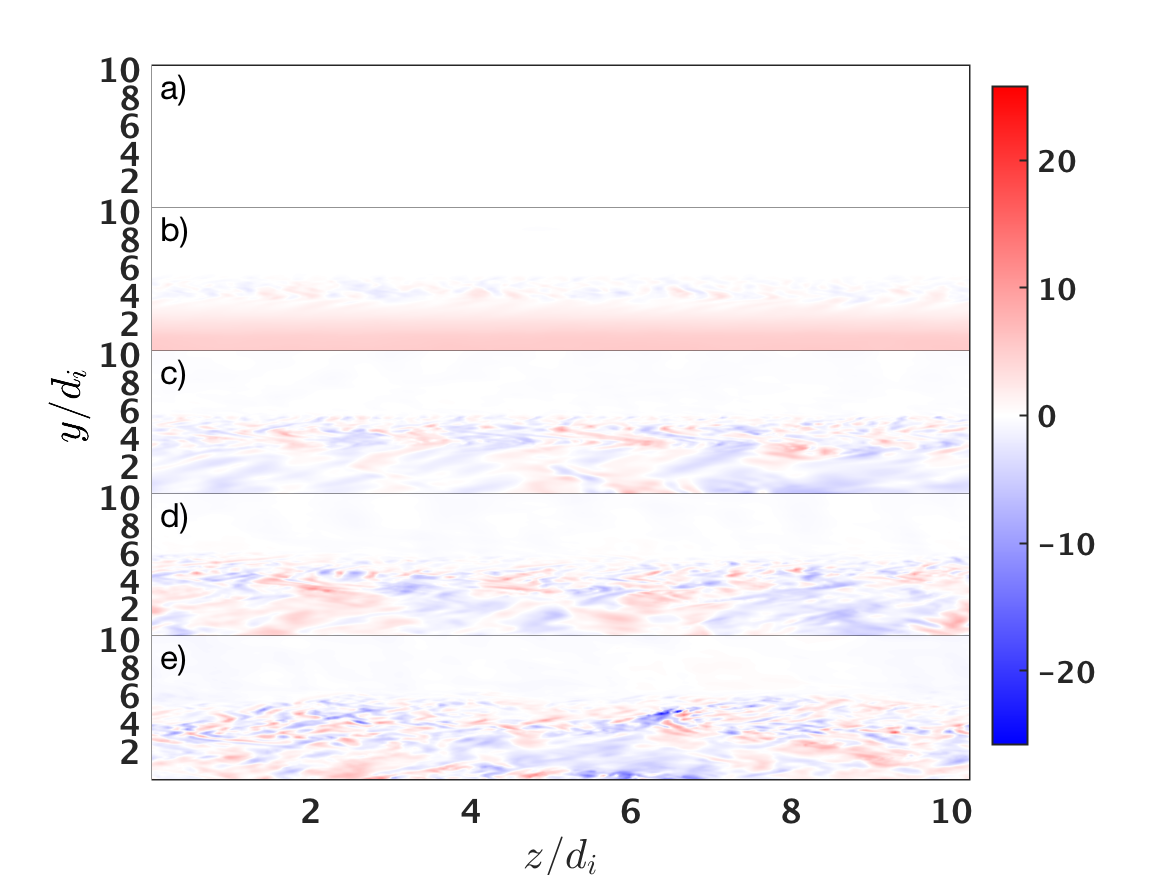}
\caption{Slices of the electron density (left) and $E_z$ (right) taken at the X-line, $x=30d_i$, for run 3D5 at times $t \Omega_{ci} = 0$ to $20$.}\label{fig:5MneEz}
\end{figure}

The origin of the turbulence in this simulation is the LHDI, in stark contrast to the ten-moment results. Evidence for this can be seen in figure~\ref{fig:5MneEz}, where the evolution of the electron density (left) and $E_z$ (right) are plotted in the $yz$-plane at times $t \Omega_{ci} = 0$ to $20$. Already, at $t \Omega_{ci} = 5$, evidence of the LHDI has become apparent, with density and electric field fluctuations extending from the the edge of the central current layer into the low density, magnetosphere, side with wavelength $k \rho_e \sim 1$. By $t \Omega_{ci} = 10$, the initially electrostatic modes have fully penetrated the current layer and fill the entire magnetosphere side. 

These results concerning the LHDI are also supported by linear stability analysis of the five-moment model presented in \citet{Ng:2019}. \citet{Ng:2019} demonstrates that the LHDI grows by a factor of two to three faster in the five-moment model than in kinetic theory, thus leading to the explosive growth of the instability and subsequent turbulence seen in run 3D5.

The generalized Ohm's for the five-moment model differs from that for ten-moment model given in equation~\eqref{eq:OhmsLaw} only in that the the off-diagonal pressure tensor components do not contribute. The one dimensional components of the generalized Ohm's law across the current layer for run 3D5 at time $t \Omega_{ci} = 20$ and $z=0$ are plotted in the right panel of figure~\ref{fig:OhmsLaw}, where the vertical dashed line indicates the location of the magnetic field reversal. The primary contributor to the reconnection electric field at the X-line is the Hall term (dashed green). Away from the X-line, the electron pressure gradient (lavender) is the primary contributor.

\section{Discussion}\label{sec:discussion}
The ten-moment model presented in section~\ref{sec:modelsTen} represents the most sophisticated fluid closure available that is numerically tractable for global magnetosphere modeling. The closure employed by the ten-moment model captures many aspects of kinetic theory that are well beyond the capabilities of the most commonly applied model to global magnetospheres, MHD. MHD assumes the plasma is highly collisional, such that it is in local thermodynamic equilibrium; therefore, the only mechanism available to break field lines and allow magnetic reconnection to proceed is resistivity, be it grid-scale diffusion or explicitly imposed. In either case, the resistivity employed exceeds the real value in weakly collisional planetary magnetospheres by orders of magnitude, thus altering significantly the location, evolution, and transport induced by reconnection. Indeed, MHD models predict either extended Sweet-Parker layers or copious plasmoid production, with the evolution being determined entirely by the Lundquist number rather than kinetic processes \citep{Daughton:2012}. Additionally, although the ten-moment model has difficulty capturing the LHDI, the LHDI is completely ordered out of MHD.

Fluid models more sophisticated than MHD exist, such as Hall MHD or two-fluid models with isothermal closures, but they are rarely employed in the context of global modelling due to computational constraints. However, both Hall MHD and traditional two-fluid models also neglect the electron kinetic effects that permit reconnection to proceed naturally, relying on resistivity to break field lines. As demonstrated by the five-moment model in section~\ref{sec:resultsFive}, retaining finite electron inertia effects without fully evolving the pressure tensor can lead to anomalously large growth rates for instabilities that are common to global magnetospheres, such as the LHDI. 

As computational resources are expanding, hybrid approaches are gaining increased focus, because they treat the ions kinetically; however, hybrid models suffer many of the same issues outlined above. Since the electrons are treated as a fluid, they also require a closure model, which is that of a massless, isothermal fluid, although finite electron mass is occasionally retained. Therefore, reconnection still proceeds via resistivity, despite having fully kinetic ions. The electron closure also feeds back on the ions through Maxwell's equations, leading to consequences at large-scales, such as undamped fast magnetosonic modes \citep{Groselj:2017} and electron equation of state dependent energy transport \citep{Parashar:2014}. Further, electron instabilities such as the LHDI remain problematic in hybrid models, because they occur on electron kinetic scales and often involve electron phase space resonances. 

Despite the overall success of the application of the ten-moment model to the Burch event presented in Section~\ref{sec:resultsTen}, it does display some weaknesses relative to fully kinetic, particle-in-cell simulations. Specifically, the ten-moment closure fails to properly capture the LHDI and the important role it plays in heating and transporting plasma across the separatrix. The source of the disagreement follows directly from the local closure chosen in section~\ref{sec:modelsTen}. \citet{Ng:2019} demonstrates that a non-local heat flux closure is able to accurately recreate the LHDI, where the local heat flux closure presented in equation~\eqref{eq:heatClosure} is replaced with
\begin{equation}
    \hat{q}_{ijk} = -i \frac{v_t}{|k|} \chi k_{[i}\hat{T}_{jk]},
\end{equation}
where $Q_{ijk} = n \tilde{q}_{ijk}$, $\hat{*}$ represents the Fourier transform, and $\chi$ is an adjustable constant. The $|k|$ factor appearing in the denominator makes this closure non-local in real space, drastically increasing the computational expense and making its utility for global modelling limited. A gradient-based closure to the ten-moment model has been introduced by \citet{Allmann:2018}, where equation~\eqref{eq:heatClosure} is replaced with
\begin{equation}
    \frac{\partial Q_{ijk}}{\partial x_k} = -\frac{v_t}{|k_0|} \nabla^2 (P_{ij} - p\delta_{ij}),
\end{equation}
and $k_0$ is an adjustable constant. \citet{Allmann:2018} find that the gradient closure better agrees with Vlasov and PIC simulations than the local closure presented in equation~\eqref{eq:heatClosure}, with minimal additional computational expense for typical choices of $k_0$. Therefore, the gradient closure appears to be a good alternative to the local closure for global modelling, but further study to determine how well the gradient closure captures the LHDI is necessary.

\section{Summary and Conclusions}\label{sec:conclusions}
We have performed an event study of the \citet{Burch:2016} 16 October 2015 MMS diffusion region crossing using the two-fluid, multi-moment models contained in the \gkeyll~framework. The ten-moment model evolves the full pressure tensor and is closed with a local Hammett-Perkins model for the heat flux divergence. The closure is designed to capture linear Landau damping that sets in at a fixed, isotropic, and species dependent wavenumber, $k_{0s}$. The five-moment model achieves closure by truncating the moment hierarchy at the isotropic pressure for each species, setting all higher moments to zero. The ten-moment model was applied to the Burch event in two and three dimensions, while the five-moment model was applied only in three dimensions. 

The ten-moment model was found to accurately capture aspects of the Burch event that are beyond the capabilities of conventional fluid approaches, such as MHD, Hall MHD, or two-fluid models with isothermal closures. Inclusion of the full electron pressure tensor, electron inertia, and Hall terms permits the electrons to demagnetize naturally at the X-line. The pressure agyrotropy structure and amplitude developed by the ten-moment model is consistent with existing PIC simulations of reconnection. The ten-moment model also accurately reproduces the same magnitude of electron temperature anisotropy as observed \textit{in situ} with MMS. However, the spatial location of the temperature anisotropy is on the magnetosheath side of the X-line; whereas, \textit{in situ} observations \citep{Burch:2016} and PIC simulations \citep{Le:2017} find it to be on the low density, magnetosphere side. The discrepancy is due to lower hybrid drift instability (LHDI) induced heating and transport in the PIC simulation. The chosen closure parameter for the ion ten-moment model, $k_{0i} d_i = 1/10$, is too far in scale from the fastest growing LHDI mode and is thus anomalously damped. Choosing a larger value for the closure parameter, $k_{0i} d_e \sim 1$, can reproduce the growth of the LHDI accurately but would prevent reconnection from proceeding correctly, since the electron pressure would be scattered toward isotropy at a much faster rate, precluding electron agyrotropy from forming. Electron shear flow and kink instabilities do develop in the layer in three dimensional ten-moment \gkeyll~simulation, which are also observed in PIC simulations \citep{Le:2017}. The shear flow instability generates significant turbulence and increases the transport across the current layer relative to the two dimensional \gkeyll~simulations of the Burch event.

In contrast to the ten-moment model, the LHDI grows more rapidly in the five-moment model than predicted by kinetic theory. The explosive growth of the instability and concomitant turbulence rapidly destroy the current layer and drives energy to the grid scale, where it is diffused away and lost from the simulation. Due to the explosive growth of the LHDI, the five-moment model was found to poorly model the Burch event. 

The ten-moment model in \gkeyll~is a simple but effective closure for capturing reconnection dynamics, which are fundamental for understanding the coupled sun-Earth system and other planetary magnetospheres, both within and without the solar system. The model well captures many of the kinetic aspects of the Burch event, demonstrating the efficacy of the ten-moment model for performing global magnetosphere simulations. However, the local, isotropic closure does anomalously damp small scale modes that play an important role in mixing and heating the plasma at the magnetopause. Therefore, additional improvements in the closure are possible, such as the gradient based closure introduced by \citet{Allmann:2018} and discussed in section~\ref{sec:discussion}.  These improvements are necessary to attain maximal fidelity of global fluid models of planetary magnetospheres and will be explored in future works.



\acknowledgments
The authors are grateful for fruitful discussions with Marc Swisdak, Ari Le, Mike Shay, and Paul Cassak. This research was supported by NSF grants AGS-1338944 and AGS-1622306. This research also used resources of the National Energy Research Scientific Computing Center (NERSC), a U.S. Department of Energy Office of Science User Facility operated under Contract No. DE-AC02-05CH11231. Simulation data used in this manuscript is available upon request.


\end{document}